\documentclass[11pt]{article}
\usepackage[T1]{fontenc}
\usepackage{amssymb,amsmath,amsfonts,dsfont,amsthm,bm}
\usepackage{float,rotfloat} 
\usepackage{mathrsfs}
\usepackage{multirow}
\usepackage{enumerate}
\usepackage{enumitem}
\usepackage{mathtools}
\usepackage{hyperref,url}
\usepackage{graphicx}
\usepackage[hang,small,bf]{caption}
\usepackage[top=1in, bottom=1in, left=1.0in, right=1.0 in]{geometry}
\usepackage{longtable}
\usepackage{color,xcolor}
\usepackage[round,sort,comma]{natbib}

\usepackage[title]{appendix}
\usepackage{etoolbox} 
\usepackage[english]{babel}
\usepackage[autostyle, english = american]{csquotes}
\MakeOuterQuote{"}

\patchcmd{\appendices}{\quad}{: }{}{} 

\newtheorem{thm}{Theorem}
\newtheorem{defn}{Definition}
\numberwithin{thm}{section}
\numberwithin{defn}{section}
\numberwithin{equation}{section}
\numberwithin{figure}{section}
\numberwithin{table}{section}
\newtheorem{note}{Note}[section]
\newcommand{\ID}{\mathds{1}}

\newcommand{\blue}{\color[rgb]{0,0,1}}
\definecolor{darkblue}{rgb}{0,0,0.55}

\allowdisplaybreaks

\hypersetup{,colorlinks=true,allcolors=darkblue}

\begin{document}
	
\baselineskip 5mm

\thispagestyle{empty}

\begin{center}
{\LARGE Robust Estimation of Loss Models for \\[10pt]
Lognormal Insurance Payment Severity Data}

\vspace{15mm}

{\large
Chudamani Poudyal, Ph.D. \\
{\sc\blue E-mail\/}: 
\url{chudapsw@gmail.com}
}

\vspace{20mm}

\copyright \
Copyright of this Manuscript is held by the Author!

\end{center}

\vspace{1.00in}

\begin{center}
{\sc Abstract}
\end{center}

\begin{quote}
The primary objective of this 
scholarly work is to develop 
two estimation procedures -- 
{\em maximum likelihood estimator\/} (MLE) and 
method of {\em trimmed moments\/} (MTM) --
for the mean and variance of lognormal 
insurance payment severity 
data sets affected by different loss control
mechanism, for example,
truncation 
(due to deductibles), 
censoring (due to policy limits), 
and scaling (due to coinsurance 
proportions),
in insurance and financial industries.
Maximum likelihood estimating equations for both 
payment-per-payment and payment-per-loss data sets
are derived which can be solved readily by any 
existing iterative numerical methods. 
The asymptotic distributions of those estimators
are established via Fisher information matrices. 
Further, with a goal of balancing efficiency and
robustness and to remove point masses at 
certain data points, we develop a dynamic 
MTM estimation procedures for lognormal 
claim severity models for the
above-mentioned transformed data scenarios.
The asymptotic distributional properties and 
the comparison with the corresponding MLEs of 
those MTM estimators are established along with 
extensive simulation studies.
Purely for illustrative purpose,
numerical examples for 1500 US indemnity losses
are provided which illustrate the practical
performance of the established results in this paper. 

\vspace{4mm}

{\bf\em Keywords \& Phrases\/}. 
Dynamic Estimation;
Lognormal Insurance Severity; 
Maximum Likelihood Estimators;
Trimmed Moments;
Insurance Payments; 
Loss Models; 
Robust Estimation; 
Truncated and Censored Data.
\end{quote}

\newpage

\baselineskip 8mm

\setcounter{page}{2}

\section{Introduction}
\label{sec:Introduction}

The research leading to the results of this 
work is basically motivated to find and compare
two estimation approaches for two types of insurance 
benefit payment severity data with the assumption of 
stand-alone continuous underlying loss distributions. 
In practice, due to commonly used loss control mechanism
in the 
financial and insurance industries \citep{EPUS,MR3890025},
the random variables we observe and wish to model are affected 
by data truncation (due to deductibles), censoring 
(due to policy limits), and scaling (due to coinsurance factor).
Therefore, the main focus of this paper is to develop  
{\em maximum likelihood} and a dynamic {\em method of trimmed moments}
estimators, respectively, called MLE and MTM for short,
of location and scale parameters of lognormal
insurance benefit payment claims severity. 

In the current practice, statistical inference for 
loss models is almost exclusively likelihood based
\citep[see, e.g.,][]{MR3890025},
which typically result in 
sensitive loss severity models if there is a small perturbation 
in the underlying assumed model or if the observed sample is coming 
from a contaminated distribution, \cite{MR0120720}. 
Further, following the work by 
\cite{MR2188658}, the contemporary actuarial loss modeling
literature has been directed toward composite loss modeling 
\citep[see, e.g.,][]{MR3324052,MR3807461,MR2347211,MR2969968}
which is also referred to as 
splicing 
\citep[see, e.g.,][]{MR3890025}
including mixtures of Erlangs 
\citep[see, e.g.,][]{MR4163087,MR3724936,MR3394072}.
The composite models appear to capture the tail behavior 
of the underlying distribution more accurately
\cite[see, e.g.,][and the references therein]{MR3999507}.
But as mentioned by \cite{MR3836650},
often the simple closed-form expressions 
for the pdf of composite models are not available
which could lead to more complexity of deriving
analytic sampling distributional results.
Therefore, beside many ideas from the mainstream robust
statistics literature 
\cite[see, e.g.,][]{MR0362657,MR0161415,MR2488795},
actuaries have to deal with heavy-tailed and skewed distributions, 
data truncation and censoring, identification and recycling of 
outliers, and aggregate loss, etc. 
Thus, it is appealing to search some estimation procedures
which directly work with those mentioned loss control mechanism,
are insensitive against small perturbations from the assumed models,
computationally efficient, and still reasonably balance the asymptotic
efficiency-robustness trade-offs with respect to the corresponding MLE.

As a member of log-location-scale family, 
\label{pageRef_WhyLN}
the lognormal distribution has diverse 
applications in actuarial science, business, and economics
\citep[see, e.g.,][and the references therein]{MR1987777}
which closely approximates certain types of homogeneous 
actuarial loss data 
\citep{hl79,MR3836650}.
Further, it has been established that,
even for the heterogeneous actuarial losses,
lognormal distribution is able to capture the nature
of the data set either on the head or on the tail or
on both head and tail parts of different composite models,
see, for example, 
\cite{MR2188658},
\cite{MR3474025},
\cite{MR3543061},
\cite{MR3836650},
\cite{MR3896968}, and
\cite{mmm20}.
More comprehensive investigation of 256 different 
composite models have been analyzed by
\cite{MR3999507}.

On the other hand, MTM
approach can be viewed as a special case of $L$-statistics
\citep{MR0203874} and has established itself as sufficiently general
and flexible in order to balance between estimator's efficiency and 
robustness for fitting of continuous parametric models based on 
complete ground-up loss severity data
\citep[see, e.g.,][]{MR2497558,MR3758788}.
But it is yet open to investigate the MTM performance beyond the
complete data scenario and this paper will address this issue 
for lognormal payment-per-payment and payment-per-loss 
data scenarios.
If a truncated (both singly and doubly) normal sample data set is
available then the MLE procedures for such data have been 
developed by \cite{MR0038041} and the method of moments 
estimators can be found in \cite{MR0045361} and \cite{MR0196848}.
A novel robust estimation procedure called 
{\em method of truncated moments} (MTuM) for modeling
different actuarial loss data scenarios is
initially purposed by \cite{MR3864903} and further 
developed by \cite{MR4192140}.
But the goal and motivation of this research work is different.
That is, instead of fitting composite models or designing MTuM, 
both MLE and MTM estimation procedures will be derived
for stand-alone lognormal insurance benefit payment data scenarios
with an assumption that a close fit in one tail or both tails of
the distribution is not desired but at the same time we want 
to remove partial point masses accumulated at the truncated 
and/or censored data points. 
As the main contribution of this parer,
asymptotic distributions, such as normality and consistency,
along with asymptotic relative efficiency (ARE) of those estimators  
(under transformed data) with respect to the corresponding MLEs 
are established including extensive validating simulation studies.
The developed methodologies in this paper are implemented
with real 1500 US indemnity loss data set.
It will be shown that, when properly redesigned, 
a dynamic MTM can be a robust and computationally 
efficient alternative to the MLE-based inference 
for claim severity models that are affected 
by deductibles, policy limits, and coinsurance.
Some of the technical challenges in implementing the developed 
estimation procedures in practical data analysis are discussed
along with the corresponding simplifying assumptions.

The remainder of the paper is organized as follows. 
In Section \ref{sec:Insurance_Payments}, we describe 
the two types of insurance benefit payment lognormal
random variables.
In Section \ref{sec:MLE}, the MLE procedures is developed
for those two payment type of data sets with their corresponding
asymptotic distributional properties. 
Section \ref{sec:MTM} is the corresponding section to develop 
dynamic MTM procedures for the location and scale parameters 
for those two payment models. 
Comparison of ARE of those
MTM estimators with respect to MLE are also presented. 
Section \ref{sec:SpecialCases} is for some special cases 
which are more common in operational risk modeling with 
some theoretical results.
Section \ref{sec:SimStudy} summarizes a detail simulation study.
Numerical examples to observe the performance of the 
developed estimation procedures can be found in 
Section \ref{sec:numericalExamples}.
Finally, summary and concluding remarks are offered in 
Section \ref{sec:Conclusion}.

\section{Insurance Payments}
\label{sec:Insurance_Payments}

Consider a ground-up lognormal loss random variable 
$W \sim LN(w_{0},\theta,\sigma)$ with cdf
\begin{align}
\label{eqn:LNCDF1}
F_{W}(w) 
& = 
\Phi 
\left( 
\dfrac{\log{(w-w_{0}})-\theta}
{\sigma}
\right),
\quad 
w > w_{0}, \ 
-\infty < \theta < \infty, \ 
\sigma > 0,
\end{align}
where $\theta$ and $\sigma$ are unknown parameters
to be estimated, $w_{0} > 0$ is assumed to be known, 
and $\Phi$ is the cdf of the standard normal distribution.
Clearly
$X := \log{(W-w_{0})}\sim N(\theta,\sigma^2)$ 
with cdf and pdf, respectively, given by:
\begin{align}
\label{eqn:LNCDF2}
F_{X}(x) 
= 
\Phi\left(\frac{x-\theta}{\sigma}\right) 
\quad \mbox{and} \quad 
f_{X}(x) 
= 
\frac{1}{\sigma}
\phi\left(\frac{x-\theta}{\sigma}\right),
\qquad 
-\infty < x < \infty,  
\end{align}
where 
$
\phi(x) 
= 
\left(
1/\sqrt{2\pi} \,
\right) 
e^{-x^2/2}, \ 
x \in \mathbb{R}
$
is the pdf of the standard normal distribution.
The corresponding quantile function 
$F_{X}^{-1} : (0,1) \to \mathbb{R}$ is given by 
$
F_{X}^{-1}(v) = \theta + \sigma \Phi^{-1}(v).
$

Insurance contracts have coverage modifications that need 
to be taken into 
account when modeling the underlying loss variable. 
Usually the coverage
modifications such as deductibles, policy limits, 
and coinsurance are 
introduced as loss control strategies so that unfavorable 
policyholder 
behavioral effects (e.g., adverse selection) can be minimized. 
There are 
also situations when certain features of the contract emerge naturally 
(e.g., the value of insured property in general insurance is a natural 
upper policy limit). 
Here we describe two common transformations of 
the loss variable along with the corresponding cdfs, pdfs, and qfs.

Suppose the lognormal severity random variable $W$ has ordinary 
deductible $d$, upper policy 
limit $u$, and coinsurance factor $c$ ($0 < c \leq 1$). 
These 
coverage specifications
imply that when a loss $W$ is reported, the insurance company 
is responsible for a proportion $c$ of $W$ exceeding $d$, but no more 
than $c(u-d)$. Define
\begin{equation}
\label{eqn:uToTdTotDefn1}
t := \log{(d-w_{0})}, \
T := \log{(u-w_{0})}, \
R := T-t, \ 
\gamma := \frac{t-\theta}{\sigma}, \ \mbox{and} \ 
\xi := \frac{T-\theta}{\sigma}.
\end{equation}
and note that it is possible to have $t < 0$ but $d > 0$.
Here, it is important to note that 
$d$ and $u$ refer for the lognormal 
left-truncation threshold and 
right-censored point, respectively.
On the other hand,
$t$ and $T$ are, respectively,
the corresponding normal form of 
left-truncation threshold and right
censored point as defined in 
(\ref{eqn:uToTdTotDefn1}).
Then, obviously
\begin{align}
\label{eqn:defTz}
\theta 
& = 
t-\sigma \gamma
\quad \mbox{and} \quad 
\xi 
= 
\gamma+\frac{R}{\sigma}.
\end{align} 
Also, let 
$
\bar{\Phi}(z)
= 
1 - \Phi(z)
$
be the standard normal survival function at 
$z \in \mathbb{R}$.
Finally, define
\begin{align}
\label{eqn:truncatedNormalZDef}
\Omega_{1} 
& :=
\frac{\phi(\gamma)}{\bar{\Phi}(\gamma)-\bar{\Phi}(\xi)}
\quad \mbox{and} \quad 
\Omega_{2} 
:= 
\frac{\phi(\xi)}{\bar{\Phi}(\gamma)-\bar{\Phi}(\xi)}.
\end{align}

Now, if the loss severity $W$ below the deductible $d$ is completely 
unobservable (even its frequency is unknown), then the 
{\em left-truncated\/}, {\em right-censored\/}, and 
{\em linearly transformed\/} (also known as 
{\em  payment-per-payment\/} variable) form of $W$ is defined as:
\begin{equation}
\label{p1data}
Y_{w} 
: =
c
\left(
\min\big\{W,u \big\}-d 
\right)
\, | \, W > d
= 
\left\{ 
\begin{array}{ll}
\mbox{undefined},
& W \le d; \\
c \left( W-d \right), & d < W < u; \\
c \left( u-d \right), & u \le W. \\
\end{array}
\right.
\end{equation}
The corresponding normal form of $Y_{w}$ is given by
\begin{equation}
\label{p1Ndata}
Y
: =
c\log
{
\left( 
\dfrac{Y_{w}}{c(d-w_{0})}
+ 1
\right)
}
=
c
\left(
\min\big\{ X,T \big\}-t
\right)
\, | \, X > t
=
\left\{ 
\begin{array}{ll}
\mbox{undefined}, 
& X \le t; \\
c \left( X-t \right), & t < X < T; \\
c \left( T-t \right), & T \le X. \\
\end{array}
\right.
\end{equation}
The cdf $F_{Y}$, pdf $f_{Y}$, and qf $F_{Y}^{-1}$ 
of the payment-per-payment random variable $Y$ are given by:
\begin{equation}
F_{Y}( y; \, c, t, T ) 
= 
\mbox{\bf P} 
\left[ c \left( 
\min\big\{ X, \, T \big\} - t \right) 
\leq y \, \big | \, X > t  \right] 
= 
\left\{ 
\begin{array}{ll}
0, & y \leq 0; \\[0.5ex] 
\frac{F_{X}(y/c+t) 
	- F_{X}(t)}{1-F_{X}(t)}, & 0 < y < c(T-t); \\[0.5ex]
1,  & y \geq c(T-t), \\
\end{array}
\right.
\label{p1cdf}
\end{equation}
\begin{equation}
\label{p1pdf}
f_{Y}( y; \, c, t, T ) 
= 
\left\{
\begin{array}{ll}
\frac{f_{X}(y/c+t)}{c [1-F_{X}(t)] }, & 0 < y < c(T-t); \\[1ex]
\frac{1-F_{X}(T^-)}{1-F_{X}(t)}, & y = c(T-t); \\[0.75ex]
0, & \mbox{elsewhere}, \\
\end{array}
\right.
\end{equation}
and
\begin{equation}
\label{p1qf}
F_{Y}^{-1}( v; \, c, t, T )
= 
\left\{ 
\begin{array}{ll}
c \left[ F_{X}^{-1} \big( v + (1-v) F_{X}(t) \big) - t \right], & 
0 \leq v < \frac{F_{X}(T)-F_{X}(t)}{1-F_{X}(t)}; \\[0.75ex]
c(T-t), & \frac{F_{X}(T)-F_{X}(t)}{1-F_{X}(t)} \leq v \leq 1. \\
\end{array}
\right.
\end{equation}

The scenario that no information is available about $W$ below $d$ is 
likely to occur when modeling is done based on the data acquired from 
a third party (e.g., data vendor). 
For payment data collected in house,
the information about the number of policies that did not report claims
(equivalently, resulted in a payment of 0) would be available. 
This minor modification yields different payment variable, say $Z_{w}$, 
which can be treated as {\em interval-censored\/} 
and {\em linearly transformed\/} $W$ (also called {\em payment-per-loss\/} 
random variable):
\begin{equation}
\label{p2data}
Z_{w}
:= 
c \left( \min \big\{ W, u \big\} - \min \big\{ W, d \big\} \right) 
= 
\left\{ 
\begin{array}{ll}
0, & W \leq d; \\
c \left( W-d \right), & d < W < u; \\
c \left( u-d \right), & u \leq W. \\
\end{array}
\right.
\end{equation}
The corresponding normal form of $Z_{w}$ is given by:
\begin{equation}
\label{p2NZdata}
Z
: =
c\log
{
\left( 
\dfrac{Z_{w}}{c(d-w_{0})}
+ 1
\right)
}
=
c
\left(
\min\big\{ X,T \big\}
-
\min\big\{ X,t \big\}
\right)
=
\left\{ 
\begin{array}{ll}
0, 
& X \le t; \\
c \left( X-t \right), & t < X < T; \\
c \left( T-t \right), & T \le X. \\
\end{array}
\right.
\end{equation}
The cdf $F_{Z}$, pdf $f_{Z}$, and qf $F_{Z}^{-1}$ are related 
to $F_{X}$, $f_{X}$, and $F_{X}^{-1}$ and given by:
\begin{equation}
F_{Z}( z; \, c, t, T ) 
= 
\mbox{\bf P} 
\left[ c 
{\small 
\left( \min \big\{ X, T \big\} 
- 
\min \big\{ X, t \big\} 
\right)
}
\leq z \right]
= 
\left\{ 
\begin{array}{ll}
0, & z < 0; \\[0.25ex]
F_{X}(z/c+t), & 0 \leq z < c(T-t); \\[0.25ex]
1,  & z \geq c(T-t), \\
\end{array}
\right.
\label{p2cdf}
\end{equation}
\begin{equation}
\label{p2pdf}
f_{Z}( z; \, c, t, T ) 
=  
\left\{
\begin{array}{ll}
F_{X}(t), & z = 0; \\[0.25ex]
f_{X}(z/c+t)/c, & 0 < z < c(T-t); \\[0.25ex]
1 - F_{X}(T^-), & z = c(T-t); \\[0.25ex]
0, & \mbox{elsewhere}, \\
\end{array}
\right.
\end{equation}
and
\begin{equation}
\label{p2qf}
F_{Z}^{-1}( v; \, c, t, T )
= 
\left\{ 
\begin{array}{ll}
0, & 0 \leq v \leq F_{X}(t); \\[0.25ex]
c \left( F_{X}^{-1} (v) - t \right), &
F_{X}(t) < v < F_{X}(T); \\[0.25ex]
c(T-t), & F_{X}(T) \leq v \leq 1. \\
\end{array}
\right.
\end{equation}

\begin{note}
The numbers $t$ and $T$ can be treated as deductible 
and policy limit, respectively,
for the normal random variable 
$X \sim N(\theta, \sigma^2)$ with a possibility 
of $t < 0$.
\qed 
\end{note}

\begin{note} 
The variable $\gamma$, defined in (\ref{eqn:uToTdTotDefn1}),
is considered to be the independent 
parameter of location, $\theta$, for MLE estimation purpose. 
Then, the mean, $\theta$, 
is a linear function of $\gamma$ given by (\ref{eqn:defTz}).
\qed 
\end{note}

\section{MLE} 
\label{sec:MLE}

Here, we develop MLE estimation procedures for $\theta$
and $\sigma$ under the transformed data scenarios 
given by (\ref{p1Ndata}) and (\ref{p2NZdata}).

\subsection{Payments {\em Y}}

Let $y_{1},\ldots,y_{n}$ be an {\em i.i.d.\/} sample 
given by pdf (\ref{p1pdf}) with policy limit $T$, deductible $t$, 
and coinsurance factor $c$. 
Then, with 
$
n_{1} 
:=
\sum_{i=1}^{n}\ID\{0<y_{i}<cR\} \ \mbox{and} \
n_{2} 
:= 
\sum_{i=1}^{n}\ID\{y_{i}=cR\}
$,
the corresponding log-likelihood function becomes
\begin{align}
\label{eqn:PPYMLELogLik1}
l_{y} (\gamma,\sigma) 
& := 
-\frac{n_{1}}{2}\log{(2\pi)}
+ n_{2}
\log{(\bar{\Phi}(\xi))} 
- n \log{(\bar{\Phi}(\gamma))} 
- n_{1}\log{(\sigma)}  
-\frac{1}{2}\sum_{0<y_{i}<cR}
\left(\gamma+\frac{y_{i}}{c\sigma}\right)^{2}.
\end{align}
Thus, setting 
$
\left(
\frac{\partial l_{y}}{\partial \gamma}, 
\frac{\partial l_{y}}{\partial \sigma} 
\right)
=
(0,0),
$
yields the system of MLE equations:
\begin{align}
\label{eqn:PPNormalMLEeqn1}
\left\{
\begin{array}{lll}
\displaystyle n\frac{\phi(\gamma)}{\bar{\Phi}(\gamma)}-n_{2}
\frac{\phi(\xi)}{\bar{\Phi}(\xi)}
-\sum_{0<y_{i}<cR}
\left(\gamma+\frac{y_{i}}{c\sigma}\right) 
& = & 0, \\[15pt] 
\displaystyle n_{2}\frac{R\phi(\xi)}{\sigma^2 \bar{\Phi}(\xi)} 
- \frac{n_{1}}{\sigma} 
+ \frac{1}{c\sigma^2}\sum_{0<y_{i}<cR}y_{i}\left(\gamma+\frac{y_{i}}{c\sigma}\right) & = & 0.
\end{array} \right.				
\end{align}
Define
\begin{align}
\label{eqn:PPNormalQDef}
\Omega_{y,1} 
& :=
\frac{n}{n_{1}} \frac{\phi(\gamma)}{\bar{\Phi}(\gamma)}
\quad \mbox{and} \quad 
\Omega_{y,2} 
:=
\frac{n_{2}}{n_{1}} \frac{\phi(\xi)}{\bar{\Phi}(\xi)}, 
\end{align}
then the system of MLE Equations (\ref{eqn:PPNormalMLEeqn1}) takes the form
\begin{align}
\label{eqn:PPNormalMLEeqn2}
\left\{
\begin{array}{rrr}
\sigma\left(\Omega_{y,1}-\Omega_{y,2}-\gamma\right) - c^{-1}\widehat{\mu}_{y,1} & = & 0, \\[10pt]
\sigma^{2}\left(1-\gamma(\Omega_{y,1}-\Omega_{y,2}-\gamma)
-\frac{\Omega_{y,2}R}{\sigma}\right) -  c^{-2}\widehat{\mu}_{y,2} 
& = & 0,
\end{array} \right.
\end{align}
where 
$\widehat{\mu}_{y,1}$ and $\widehat{\mu}_{y,2}$ are the first and second sample moments, 
$\widehat{\mu}_{y,j} 
:=
n_{1}^{-1}\sum_{i=1}^{n}\ID\{0<y_{i}<cR\}y_{i}^{j}$, $j = 1,2$. 
To solve the system (\ref{eqn:PPNormalMLEeqn2}) for $\widehat{\gamma}_{\mbox{\tiny y,MLE}}$ and
$\widehat{\sigma}_{\mbox{\tiny y,MLE}}$, 
we initialize the system as below: 
\begin{align}
\label{eqn:PPY_MLE_StartVal}
\sigma_{\text{start}} 
& =
\sqrt{c^{-2}\widehat{\mu}_{y,2} - \left(c^{-1}\widehat{\mu}_{y,1}\right)^{2}} 
\qquad \text{and} \qquad 
\gamma_{\text{start}} 
= 
- \frac{c^{-1}\widehat{\mu}_{y,1}}{\sigma_{\text{start}}}.
\end{align}

\begin{thm}
\label{thm:PPY_Fisher_Info}
Let
$
\bm{I}_{y}(\gamma,\sigma) 
= 
\begin{bmatrix}
a_{11} & a_{12} \\
a_{21} & a_{22} 
\end{bmatrix}
$
be the Fisher information matrix for a sample of size $1$
for the random variable $Y$, then
\[
a_{11}
= 
-\Lambda r_{1}(\gamma,\xi), \
a_{12} 
=
a_{21} 
= 
-\Lambda \sigma^{-1} r_{2}(\gamma,\xi), 
\ \mbox{and} \
a_{22}
= 
-\Lambda \sigma^{-2} r_{3}(\gamma,\xi),
\]
where 
$\Lambda 
:=
\frac{\bar{\Phi}(\gamma)-\bar{\Phi}(\xi)}
{\bar{\Phi}(\gamma)}$ and
\begin{align}
\label{eqn:PPNormalMLEFishergFunctions}
\left\{
\begin{array}{lll}
	r_{1}(\gamma,\xi) 
	& := & 
	-\left[1+\gamma\Omega_{1}-\xi\Omega_{2}
	-
	\frac{\phi(\gamma)}{\bar{\Phi}(\gamma)}\Omega_{1}  +\frac{\phi(\xi)}{\bar{\Phi}(\xi)}\Omega_{2}\right], \\[10pt]
	r_{2}(\gamma,\xi) 
	& := &
	\frac{R\Omega_{2}}{\sigma}
	\left[\frac{\phi(\xi)}{\bar{\Phi}(\xi)}
	-\xi\right]+\left[\Omega_{1}-\Omega_{2}-\gamma\right], \\[10pt]
	r_{3}(\gamma,\xi) 
	& := &
	\left(\frac{R}{\sigma}\right)^{2}\Omega_{2}
	\left(\xi-\frac{\phi(\xi)}{\bar{\Phi}(\xi)}\right)-\left[2 -\gamma(\Omega_{1}-\Omega_{2}-\gamma)-\frac{\Omega_{2}R}{\sigma}\right]. \\
\end{array} \right.				
\end{align}
Consequently, the asymptotic normality of 
$
\left(
\widehat{\gamma}_{\mbox{\tiny y,MLE}},
\widehat{\sigma}_{\mbox{\tiny y,MLE}}
\right)
$
is given by:
\begin{equation}
\label{eqn:MLEYDeltaMatrix0}
\sqrt{n}\left(
\widehat{\gamma}_{\mbox{\tiny y,MLE}}
-
\gamma,
\widehat{\sigma}_{\mbox{\tiny y,MLE}}
-\sigma
\right)
\stackrel{D}\longrightarrow 
N\left((0,0),
\bm{I}_{y}^{-1}(\gamma,\sigma) 
\right).
\end{equation}

\begin{proof}
First note that 
$\bm{I}_{y}(\gamma,\sigma)
=
n^{-1}\bm{I}_{y,n}(\gamma,\sigma)$,
where $\bm{I}_{y,n}(\gamma,\sigma)$
is the Fisher information matrix for a sample of 
size $n$.
Using (\ref{eqn:PPNormalMLEeqn2}), it can 
be shown that
\[
\dfrac{\partial^2 l_{y}}{\partial \gamma^{2}}
= 
n_{1}
R_{1}(\gamma,\xi), \quad 
\dfrac{\partial^2 l_{y}}{\partial \sigma \ \partial \gamma}
= 
\dfrac{n_{1}}{\sigma}
R_{2}(\gamma,\xi), \quad 
\dfrac{\partial^2 l_{y}}{\partial \sigma^{2}}
= 
\dfrac{n_{1}}{\sigma^2}
R_{3}(\gamma,\xi),
\]
where 
\begin{align}
	\label{eqn:PPNormalMLEFishergFunctionsRVs}
	\left\{
	\begin{array}{lll}
		R_{1}(\gamma,\xi) 
		& := & 
		-\left[1+\gamma \Omega_{y,1}-\xi \Omega_{y,2}
		-\frac{n_{1}}{n}\Omega_{y,1}^{2} 
		+\frac{n_{1}}{n_{2}}\Omega_{y,2}^{2}\right], \\[10pt]
		R_{2}(\gamma,\xi) 
		& := & 
		\frac{R\Omega_{y,2}}{\sigma}
		\left[\frac{n_{1}}{n_{2}}\Omega_{y,2}-\gamma\right]
		+\left[\Omega_{y,1}-\Omega_{y,2}-\gamma\right], \\[10pt]
		R_{3}(\gamma,\xi) 
		& := & 
		\left(\frac{R}{\sigma}\right)^{2}\Omega_{y,2}
		\left(\xi-\frac{n_{1}}{n_{2}}
		\frac{\phi(\xi)}{\bar{\Phi}(\xi)}\right)
		-\left[2 -\gamma(\Omega_{y,1}-\Omega_{y,2}-\gamma)-\frac{\Omega_{y,2}R}{\sigma}\right]. 
	\end{array} \right.				
\end{align}
Since $n_{1} \sim \mbox{Binomial}(n,\Lambda)$
and $n_{2} \sim \mbox{Binomial}
\left(n,\bar{\Phi}(\xi)/\bar{\Phi}(\gamma)\right)$,
then it follows that
\begin{align}
	\left\{
	\begin{array}{lll}
		a_{11}
		& = & 
		-\frac{1}{n}
		E
		\left[
		\dfrac{\partial^2 l_{y}}{\partial \gamma^{2}}
		\right] 
		=
		-\frac{1}{n}
		E\left[
		n_{1}R_{1}(\gamma,\xi)
		\right]
		= 
		-\Lambda r_{1}(\gamma,\xi), \\[12pt]
		a_{12}
		& = & 
		-\frac{1}{n}
		E
		\left[
		\dfrac{\partial^2 l_{y}}{\partial \sigma \ \partial \gamma}
		\right]
		=
		-\frac{1}{n}
		E\left[
		\frac{n_{1}}{\sigma} R_{2}(\gamma,\xi)
		\right]
		= 
		- \Lambda \sigma^{-1} 
		r_{2}(\gamma,\xi), \\[12pt]
		a_{22}
		& = & 
		-\frac{1}{n}
		E
		\left[
		\dfrac{\partial^2 l_{y}}{\partial \sigma^{2}}
		\right]
		=
		-\frac{1}{n}
		E\left[
		\frac{n_{1}}{\sigma^2} R_{3}(\gamma,\xi)
		\right]
		= 
		-\Lambda \sigma^{-2} r_{3}(\gamma,\xi),
	\end{array}
	\right.
\end{align}
as desired.
\end{proof}
\end{thm}

From Equation
\eqref{eqn:MLEYDeltaMatrix0},
it follows that
\begin{equation}
\label{eqn:MLEYDeltaMatrix1}
(\widehat{\gamma}_{\mbox{\tiny y,MLE}},
\widehat{\sigma}_{\mbox{\tiny y,MLE}}) 
\sim 
\mathcal{AN}\left((\gamma,\sigma),
\frac{\Lambda^{-1}}{n\left(r_{1}r_{3}-r_{2}^{2}\right)}
\begin{bmatrix}
-r_{3} & \sigma r_{2} \\[10pt]
\sigma r_{2} & -\sigma^{2}r_{1} 
\end{bmatrix}
\right).
\end{equation}
Further, since 
$
(\theta,\sigma)
=
(t-\sigma \gamma, \sigma),
$
then by multivariate delta method
\citep[see, e.g.,][p. 122]{MR595165},
we have
$
(\widehat{\theta}_{\mbox{\tiny y,MLE}},
\widehat{\sigma}_{\mbox{\tiny y,MLE}}) 
\sim 
\mathcal{AN}
\left(
(\theta,\sigma),
\frac{1}{n} \bm{S}_{\mbox{\tiny {y,MLE}}}
\right),
$
where 
\begin{align}
\label{eqn:MLEYDeltaMatrix2}
\bm{S}_{\mbox{\tiny {y,MLE}}}
= 
\frac{\Lambda^{-1}}{\left(r_{1}r_{3}-r_{2}^{2}\right)}
\mathbf{D}
\begin{bmatrix}
-r_{3} & \sigma r_{2} \\[10pt]
\sigma r_{2} & -\sigma^{2}r_{1} 
\end{bmatrix}
\mathbf{D}'
\quad \mbox{and} \quad 
\mathbf{D}
& =
\begin{bmatrix}
-\sigma & -\gamma \\
0 & 1
\end{bmatrix}.
\end{align}

\subsection{Payments {\em Z}}

Consider an observed {\em i.i.d.\/} sample $z_{1},\ldots,z_{n}$
given by pdf (\ref{p2pdf}).
Define
\begin{align}
\label{defn:n0n1n2PPZ}
n_{0} & := \sum_{i=1}^{n}\ID\{z_{i}=0\}, 
& n_{1} & := \sum_{i=1}^{n}\ID\{0<z_{i}<cR\}, 
& n_{2} & := \sum_{i=1}^{n}\ID\{z_{i}=cR\}.
\end{align}
Note that $n=n_{0}+n_{1}+n_{2}$. 
In this case the log-likelihood function becomes
\begin{align}
\label{eqn:PPZMLELogLik1}
l_{z}(\gamma,\sigma)
& :=
- \frac{n_{1}}{2}\log{(2\pi)}
+ n_{0}
\log{
\left(
1-\bar{\Phi}(\xi)
\right)
} 
- n_{1}\log{(\sigma)} \nonumber \\
& {\qquad} 
- \frac{1}{2}\sum_{0<z_{i}<cR}
\left(\gamma+\frac{z_{i}}{c\sigma}\right)^{2} 
+ n_{2}
\log{
\left(
\bar{\Phi}(\xi)
\right)
}.
\end{align}
The corresponding likelihood system of equations takes the form:
\begin{align}
\label{eqn:censoredNormalMLEeqn1}
\left\{
\begin{array}{lllll}
\frac{\partial l_{z}}{\partial \gamma} 
& = & 
\displaystyle 
n_{0}\frac{\phi(\gamma)}{{\Phi}(\gamma)}-n_{2}
\frac{\phi(\xi)}{\bar{\Phi}(\xi)}
-\sum_{0<z_{i}<cR}\left(\gamma
+\frac{z_{i}}{c\sigma}\right) 
& = & 0, \\[15pt] 
\frac{\partial l_{z}}{\partial \sigma} 
& = &
\displaystyle 
-n_{1}\frac{1}{\sigma}
+\frac{1}{\sigma^2}\sum_{0 < z_{i} < cR}
\left[\frac{z_{i}}{c}\left(\gamma
+\frac{z_{i}}{c\sigma}\right)\right] 
+ n_{2}\frac{\phi(\xi)}{\bar{\Phi}(\xi)} 
& = & 0.
\end{array} \right.				
\end{align}
Similarly with (\ref{eqn:truncatedNormalZDef}), define
\begin{align}
\label{eqn:censoredNormalYDef}
\Omega_{z,1} 
& :=
\frac{n_{0}}{n_{1}}\frac{\phi(\gamma)}{{\Phi}(\gamma)}, \qquad 
\Omega_{z,2} := \frac{n_{2}}{n_{1}}{\frac{\phi(\xi)}{\bar{\Phi}(\xi)}}, 
\end{align}
then the MLE system of Equations (\ref{eqn:censoredNormalMLEeqn1}) becomes:
\begin{align}
\label{eqn:censoredNormalMLEeqn2}
\left\{
\begin{array}{rrr}
\sigma\left(\Omega_{z,1}-\Omega_{z,2}-\gamma\right) 
- c^{-1}\widehat{\mu}_{z,1} & = & 0, \\[10pt]
\sigma^{2}\left(1-\gamma(\Omega_{z,1}-\Omega_{z,2}-\gamma)-\frac{\Omega_{z,2}R}{\sigma}\right)
- c^{-2}\widehat{\mu}_{z,2} & = & 0,
\end{array} \right.				
\end{align}
where 
$\widehat{\mu}_{z,1}$ and
$\widehat{\mu}_{z,2}$ 
are the first and second sample moments, 
$\widehat{\mu}_{z,j} 
:= 
n_{1}^{-1}\sum_{i=1}^{n}\ID\{0<z_{i}<cR\}z_{i}^{j}$, $j=1,2.
$
With the assumptions of $n_{0} > 0$ and $n_{2} > 0$,
the system (\ref{eqn:censoredNormalMLEeqn2}) with 
relation (\ref{eqn:defTz}) can be initialized at 
\citep{MR0038041}:
\[
\left( 
\gamma_{\mbox{\tiny start}}, 
\sigma_{\mbox{\tiny start}} 
\right)
=
\left(
\Phi^{-1}
\left(
\frac{n_{0}}{n}
\right),
\frac{R}
{\xi_{\text{start}}-\gamma_{\text{start}}}
\right),
\quad \mbox{where} \quad 
\xi_{\mbox{\tiny start}}
=
\Phi^{-1}
\left(
1-n_{2}/n
\right).
\]
Note that 
$\gamma_{\mbox{\tiny start}}$ and
$\sigma_{\mbox{\tiny start}}$
are simply the empirically estimated values 
of $\gamma$ and $\sigma$, respectively.
And, if $n_{2} = 0$  which implies that none of the 
observations are censored, then 
\eqref{eqn:leftTruncatedUniqueSol2}
can provide a satisfactory initialization 
of the system 
\eqref{eqn:censoredNormalMLEeqn2}.

\begin{thm}
\label{thm:PLZ_Fisher_Info}
Let
$
\bm{I}_{z}(\gamma,\sigma)
= 
\begin{bmatrix}
b_{11} & b_{12} \\
b_{21} & b_{22} 
\end{bmatrix}
$
be the Fisher information matrix for a sample of size $1$
for the random variable $Z$, then
\[
b_{11}
= 
-\Lambda \bar{\Phi}(\gamma) \ \psi_{1}(\gamma,\xi), \
b_{12} 
=
b_{21} 
= 
-\Lambda \bar{\Phi}(\gamma) \ \sigma^{-1} \psi_{2}(\gamma,\xi), 
\ \mbox{and} \
b_{22}
= 
-\Lambda \bar{\Phi}(\gamma) \ \sigma^{-2} \psi_{3}(\gamma,\xi),
\]
where $\Lambda$ is defined as in Theorem
\ref{thm:PPY_Fisher_Info} and 
\begin{align}
\label{eqn:censoredNormalMLEFishergFunctions}
\left\{
\begin{array}{lll}
	\psi_{1}(\gamma,\xi) 
	& := &
	-\left[1+\gamma\Omega_{1}-\xi\Omega_{2}
	+\frac{\phi(\gamma)}{\Phi(\gamma)}\Omega_{1}  +\frac{\phi(\xi)}{\bar{\Phi}(\xi)}\Omega_{2}\right], \\[10pt]
	\psi_{2}(\gamma,\xi) 
	& := &
	\frac{R\Omega_{2}}{\sigma}\left[\frac{\phi(\xi)}{\bar{\Phi}(\xi)}-\xi\right]+\left[\Omega_{1}-\Omega_{2}-\gamma\right], \\[10pt]
	\psi_{3}(\gamma,\xi)
	& := & 
	\left(\frac{R}{\sigma}\right)^{2}\Omega_{2}\left(\xi-\frac{\phi(\xi)}{\bar{\Phi}(\xi)}\right)-\left[2 -\gamma(\Omega_{1}-\Omega_{2}-\gamma)-\frac{\Omega_{2}R}{\sigma}\right]. \\
\end{array} \right.				
\end{align}
Consequently, the asymptotic normality of 
$
\left(
\widehat{\gamma}_{\mbox{\tiny z,MLE}},
\widehat{\sigma}_{\mbox{\tiny z,MLE}}
\right)
$
is given by:
\begin{equation}
\label{eqn:MLEZDeltaMatrix0}
\sqrt{n}\left(
\widehat{\gamma}_{\mbox{\tiny z,MLE}}
-
\gamma,
\widehat{\sigma}_{\mbox{\tiny z,MLE}}
-\sigma
\right)
\stackrel{D}\longrightarrow 
N\left((0,0),
\bm{I}_{z}^{-1}(\gamma,\sigma) 
\right).
\end{equation}

\begin{proof}
Using (\ref{eqn:censoredNormalMLEeqn2}), it can 
be shown that
\[
\dfrac{\partial^2 l_{z}}{\partial \gamma^{2}}
= 
n_{1}
\Psi_{1}(\gamma,\xi), \quad 
\dfrac{\partial^2 l_{z}}{\partial \sigma \ \partial \gamma}
= 
\dfrac{n_{1}}{\sigma}
\Psi_{2}(\gamma,\xi), \quad 
\dfrac{\partial^2 l_{z}}{\partial \sigma^{2}}
= 
\dfrac{n_{1}}{\sigma^2}
\Psi_{3}(\gamma,\xi),
\]
where 
\begin{align}
	\label{eqn:censoredNormalMLEFishergFunctionsRVs}
	\left\{
	\begin{array}{lll}
		\Psi_{1}(\gamma,\xi) 
		& := & 
		-\left[1+\gamma \Omega_{z,1}-\xi \Omega_{z,2}
		+ \frac{n_{1}}{n_{0}}\Omega_{z,1}^{2}  
		+ \frac{n_{1}}{n_{2}}\Omega_{z,2}^{2}\right], \\[10pt]
		\Psi_{2}(\gamma,\xi) 
		& := & 
		\frac{R \Omega_{z,2}}{\sigma}
		\left[
		\frac{n_{1}}{n_{2}}\Omega_{z,2}-\xi
		\right]
		+\left[\Omega_{z,1}-\Omega_{z,2}-\gamma\right], \\[10pt]
		\Psi_{3}(\gamma,\xi) 
		& := & 
		\left(\frac{R}{\sigma}\right)^{2}\Omega_{z,2}
		\left(
		\xi-\frac{n_{1}}{n_{2}}\Omega_{z,2}
		\right)
		- \left[2 -\gamma(\Omega_{z,1}-\Omega_{z,2}-\gamma)-\frac{\Omega_{z,2}R}{\sigma}\right]. \\
	\end{array} \right.				
\end{align}
Also, note that 
$n_{0} \sim \mbox{Binomial}(n,1-\bar{\Phi}(\gamma))$,
$n_{1} \sim \mbox{Binomial}(n,\bar{\Phi}(\gamma)-\bar{\Phi}(\xi))$, and
$n_{2} \sim \mbox{Binomial}(n,\bar{\Phi}(\xi))$,
then it follows that
\begin{align}
	\left\{
	\begin{array}{lll}
		b_{11}
		& = & 
		-\frac{1}{n}
		E
		\left[
		\dfrac{\partial^2 l_{z}}{\partial \gamma^{2}}
		\right]
		=
		-\frac{1}{n}
		E\left[
		n_{1}\Psi_{1}(\gamma,\xi)
		\right]
		= 
		-\Lambda \bar{\Phi}(\gamma) \ \psi_{1}(\gamma,\xi), \\[12pt]
		b_{12}
		& = & 
		-\frac{1}{n}
		E
		\left[
		\dfrac{\partial^2 l_{z}}{\partial \sigma \ \partial \gamma}
		\right]
		=
		-\frac{1}{n}
		E\left[
		\frac{n_{1}}{\sigma} \Psi_{2}(\gamma,\xi)
		\right]
		= 
		-\Lambda \bar{\Phi}(\gamma) \ \sigma^{-1} \psi_{2}(\gamma,\xi), \\[12pt]
		b_{22}
		& = & 
		-\frac{1}{n}
		E
		\left[
		\dfrac{\partial^2 l_{z}}{\partial \sigma^{2}}
		\right]
		=
		-\frac{1}{n}
		E\left[
		\frac{n_{1}}{\sigma^2} \Psi_{3}(\gamma,\xi)
		\right]
		= 
		-\Lambda \bar{\Phi}(\gamma) \ \sigma^{-2} \psi_{3}(\gamma,\xi), 
	\end{array}
	\right.
\end{align}
as desired.
\end{proof}
\end{thm}

From Equation \eqref{eqn:MLEZDeltaMatrix0},
it follows that
\begin{equation}
\label{eqn:MLEZDeltaMatrix1}
(\widehat{\gamma}_{\mbox{\tiny z,MLE}},
\widehat{\sigma}_{\mbox{\tiny z,MLE}}) 
\sim \mathcal{AN}\left((\gamma,\sigma),
\frac{\Lambda^{-1}}{n \bar{\Phi}(\gamma)
\left(\psi_{1}\psi_{3}-\psi_{2}^{2}\right)}
\begin{bmatrix}
-\psi_{3} & \sigma \psi_{2} \\[10pt]
\sigma \psi_{2} & -\sigma^{2}\psi_{1} 
\end{bmatrix}
\right),
\end{equation}
consequently  
$(\widehat{\theta}_{\mbox{\tiny z,MLE}},
\widehat{\sigma}_{\mbox{\tiny z,MLE}}) 
\sim 
\mathcal{AN}
\left((\theta,\sigma),
\frac{1}{n}\bm{S}_{\mbox{\tiny {z,MLE}}}
\right),
$
where 
\begin{equation}
\label{eqn:MLEZDeltaMatrix2}
\bm{S}_{\mbox{\tiny{z,MLE}}}
=
\frac{\Lambda^{-1}}
{\bar{\Phi}(\gamma)\left(\psi_{1}\psi_{3}-\psi_{2}^{2}\right)}
\mathbf{D}
\begin{bmatrix}
-\psi_{3} & \sigma \psi_{2} \\[10pt]
\sigma \psi_{2} & -\sigma^{2}\psi_{1} 
\end{bmatrix}
\mathbf{D}',
\quad
\mathbf{D}
\ \mbox{is given by (\ref{eqn:MLEYDeltaMatrix2})}.
\end{equation}

\section{MTM}
\label{sec:MTM}

MTMs are derived by following the standard method-of-moments 
approach, but instead of standard moments we match sample and population 
{\em trimmed\/} moments (or their variants), 
see, for example, \cite{MR2497558}. 
Here, we develop MTM estimation procedures for $\theta$
and $\sigma$ under the transformed data scenarios 
given by (\ref{p1Ndata}) and (\ref{p2NZdata}).

\begin{defn}[Method of Trimmed Moments -- MTM]
Let 
$
\boldsymbol{\theta}
= 
(\theta_{1},\ldots,\theta_{k})
$
be the parameter vector to be estimated.
For the random variables defined by (\ref{p1Ndata}) or
(\ref{p2NZdata}), let us denote the sample and population trimmed
moments as $\widehat{\mu}_{j}$ and 
$\widehat{\mu}_j(\boldsymbol{\theta})$, respectively. 
Let $v_{1:n} \leq \cdots \leq v_{n:n}$ be an ordered realization of variables 
(\ref{p1Ndata}) or (\ref{p2NZdata}) 
with qf denoted $F^{-1}_V(v \, | \, \boldsymbol{\theta})$ 
where 
$
V 
\in 
\{ 
Y,Z
\},
$
then the sample and population 
trimmed moments, with the trimming proportions $a$ (lower) and
$b$ (upper), have the following expressions:
\begin{eqnarray}
\widehat{\mu}_{j} 
& = & 
\frac{1}{n-m_n-m_n^*}
\sum_{i = m_n + 1}^{n - m_n^*}
\big[ h(v_{i:n}) \big]^j, 
\qquad j = 1, \ldots, k,
\label{Ts}
\\[1ex]
\mu_j(\boldsymbol{\theta}) 
& = &
\frac{1}{1-a-b} \int_{a}^{1-b} 
\big[ h (F_V^{-1}(v \, | \, \boldsymbol{\theta})) \big]^j \, dv,
\qquad j = 1, \ldots, k.
\label{Tp}
\end{eqnarray}
The trimming proportions $a$ and $b$ 
and function $h$ are chosen by the researcher. 
The function $h$ is specially chosen for 
"mathematical convenience" depending upon 
the nature of the underlying distribution 
function, for more details, see, for example, \cite{MR2497558}.
Also, integers $m_n$ and 
$m_{n}^* ~ (0 \le m_n < n - m_n^* \le n)$ are such that $m_n/n \rightarrow a$ 
and $m_n^*/n \rightarrow b$ when $n \rightarrow \infty$. In finite samples, 
the integers $m_n$ and $m_{n}^*$ are computed as $m_n = [n a]$ and 
$m_{n}^* = [n b]$, where $[\cdot]$ denotes the greatest integer part. 

The MTM estimators are then found by matching sample trimmed moments 
(\ref{Ts}) with population trimmed moments (\ref{Tp}) for 
$j = 1, \ldots, k$, and then solving 
the system of equations with respect to 
$\theta_1, \ldots, \theta_k$.
The obtained solutions, which we denote by 
$\widehat{\theta}_j = g_j(\widehat{\mu}_{1}, \ldots, \widehat{\mu}_{k})$,
$1 \leq j \leq k$, are, by definition, the MTM estimators of 
$\theta_1, \ldots, \theta_k$. 
Note that the functions $s_j$ are such that
$\theta_j = g_j(\mu_1(\boldsymbol{\theta}), \ldots, \mu_k(\boldsymbol{\theta}))$.
\end{defn}

MTM estimators belong to the class of $L$-statistics whose general asymptotic 
properties have been established by \cite{MR0203874}. 
A computationally more efficient formulation has been derived by \cite{MR2497558} and is given by Theorem \ref{thm:PPY_MTM_AS1}.

\begin{thm}
\label{thm:PPY_MTM_AS1}
Suppose an i.i.d. realization of variables 
(\ref{p1data}) or (\ref{p2data}) 
has been generated by cdf $F_V(v \, | \, \boldsymbol{\theta})$ which 
depending upon the data scenario equals to cdf (\ref{p1cdf}) or (\ref{p2cdf}), respectively. 
Let 
$\widehat{\boldsymbol{\theta}}_{\mbox{\tiny MTM}} 
= 
\left( \widehat{\theta}_1, \ldots, \widehat{\theta}_k \right) 
= 
\left( g_1(\widehat{\mu}_{1}, \ldots, \widehat{\mu}_{k}), \ldots,
g_k(\widehat{\mu}_{1}, \ldots,\widehat{\mu}_{k}) \right)$ denote 
an MTM estimator of $\boldsymbol{\theta}$. 
Then,
\begin{align}
\label{eqn:MTM_PPY_AsymResult1}
\widehat{\boldsymbol{\theta}}_{\mbox{\tiny MTM}} 
= 
\left( \widehat{\theta}_1, \ldots, \widehat{\theta}_k \right) ~~is~~ 
{\cal{AN}}
\left( 
\big( \theta_1, \ldots, \theta_k \big), \, \frac{1}{n} \, 
\mathbf{D} \boldsymbol{\Sigma}\mathbf{D}'
\right),
\end{align}
where 
$\mathbf{D}
:=
\big[ d_{ij} \big]_{i,j=1}^{k}$ is the Jacobian 
of the transformations 
$g_1, \ldots, g_k$ evaluated at 
$\big( \mu_1(\boldsymbol{\theta}), \ldots, \mu_k(\boldsymbol{\theta}) \big)$
and $\mathbf{\Sigma}
:= 
\big[ \sigma^2_{ij} \big]_{i,j=1}^{k}$ is 
the variance-covariance matrix with the entries:
\begin{align}
\label{eq:mtmVB_var_cov}
\sigma^2_{ij} 
= 
\frac{1}{(1-a-b)^{2}}
\int_{a}^{1-b} \int_{a}^{1-b}
\big( \min \{ v, w \} - v w \big) \;
\mbox{d} \left[ h \big( F_V^{-1}(v) \big) \right]^j \,
\mbox{d} \left[ h \big( F_V^{-1}(w) \big) \right]^i.
\end{align}
\end{thm}

The asymptotic performance of the newly designed 
estimators will be measured via ARE with respect 
to MLE and for two parameter case it is defined as 
\citep[see, e.g.,][]{MR595165,MR1652247}:
\begin{equation} 
\label{eq:infinite_relative_efficiency_benchmark_MLE}
ARE(\mathcal{C}, MLE) 
=
\left( 
\dfrac{\mbox{det}
\left(\bm{\Sigma}_{\mbox{\tiny MLE}}\right)}
{\mbox{det}
\left(\bm{\Sigma}_{\mbox{\tiny $\mathcal{C}$}}\right)}
\right)^{1/2},
\end{equation}
where 
$
\bm{\Sigma}_{\mbox{\tiny MLE}}
$
and 
$
\bm{\Sigma}_{\mbox{\tiny $\mathcal{C}$}}
$
are the asymptotic variance-covariance matrices of the 
MLE and $\mathcal{C}$ estimators, respectively, 
and det stands for the determinant of a square matrix.
The main reason why MLE should be used as a benchmark is its
optimal asymptotic performance in terms of variability 
(of course,  with the usual caveat of "under certain regularity conditions"), for more details we refer to 
\cite{MR595165} \S4.1.

\subsection{Payments {\em Y}}
\label{sec:PPY_MTM1}

For payment-per-payment data, there are three different cases to
be considered. 
From the quantile function (\ref{p1qf}), define 
$
s^{*}
=
\frac{F_{X}(T)-F_{X}(t)}{1-F_{X}(t)},
$
then we get the following arrangements:

\medskip

Case 1: 
$0 \leq a < s^{*} \leq 1-b \leq 1$ 
~(estimation based on observed and censored data).

Case 2: 
$0 \leq a < 1-b \leq s^{*} \leq 1$ 
~(estimation based on observed data only).

Case 3: 
$0 < s^{*} \leq a < 1-b \leq 1$ 
~(estimation based on censored data only).

\medskip

\noindent
In all these cases, the sample trimmed moments (\ref{Ts})
can be easily computed by first empirically estimating the probability:
\begin{align}
\label{eqn:defnSStar1}
\hat{s}_{\mbox{\tiny E}}^{*}
& :=
\frac{F_{n}(T)-F_{n}(t)}{1-F_{n}(t)}
=
n^{-1} \sum_{i=1}^n \mbox{\large\bf 1} \{ 0 < y_i < cR \},
\quad \mbox{where $F_{n}$ is the empirical cdf},
\end{align}
then selecting $a$, $b$, and finally choosing 
$h_Y(y) := y/c+t$. 
Note that $c$, $t$, and $T$ are known constants.

More specifically, if both uncensored and censored sample observations
participate in  $\widehat{\mu}_{j}$, then we end up with the first case. 
If the censored observations, that is, 
$y_{i} = cR, \ 1 \leq i \leq n$ 
are not involved in computing 
$\widehat{\mu}_{j}$, then we end up with the second case. 
And, finally if $\widehat{\mu}_{j}$ is computed only with censored 
observations, then we are in the third case, but in this case 
the population trimmed moment $\mu_{j}$ is no longer a 
function of the parameter to be estimated. 
We may also rule out the third case by choosing $a_{j}=0$, 
that is, no trimming on the left which is reasonable since the 
sample data set is already left truncated at $t$.
Further, due to space limitations and to stay with the observed
data only, from now on we only focus on Case 2.

Now, let $y_{1},\ldots,y_{n}$ be an {\em i.i.d.} sample of normal 
payment-per-payment data defined by (\ref{p1Ndata})
with qf (\ref{p1qf}). 
Then,
\begin{align}
\label{eqn:PPYMTMHat1}
\left\{
\begin{array}{lcl}
\widehat{\mu}_{y,1} 
& = &
\displaystyle
\frac{1}{n-m_{n}-m_{n}^{*}}
\sum_{i=m_{n}+1}^{n-m_{n}^{*}}
h_{Y}(y_{i:n})
=
\frac{1}{n-m_{n}-m_{n}^{*}}\sum_{i=m_{n}+1}^{n-m_{n}^{*}}
\left(\frac{y_{i:n}}{c}+t\right), \\[20pt]
\widehat{\mu}_{y,2} 
& = &
\displaystyle
\frac{1}{n-m_{n}-m_{n}^{*}}
\sum_{i=m_{n}+1}^{n-m_{n}^{*}}
\left(
h_{Y}(y_{i:n})
\right)^{2}
=
\frac{1}{n-m_{n}-m_{n}^{*}}
\sum_{i=m_{n}+1}^{n-m_{n}^{*}}
{\left(\frac{y_{i:n}}{c}+t\right)^{2}},
\end{array}
\right.
\end{align}
with $m_{n}/n \rightarrow a$ and $m_{n}^{*}/n \rightarrow b$. 
With Case 2, choose
$m_{n}^{*} \geq \sum_{i=1}^{n}\ID\{y_{i} = cR\}$.
The corresponding population trimmed moments  
(\ref{Tp}) with the qf defined by (\ref{p1qf}) are given by:
\begin{align}
\mu_{y,1}  
& = 
\frac{1}{1-a-b} 
\int_{a}^{1-b}
h_{Y}
\left(
F_{Y}^{-1}(s) 
\right)\, ds \nonumber \\
& =
\frac{1}{1-a-b} \int_{a}^{1-b}\left[ F_{X}^{-1} \big( s + (1-s) F_{X}(t \, | \, \bm{\theta})|\bm{\theta}\big) \right] \, ds , 
\nonumber \\
& = 
\theta +\sigma c_{y, 1}, 
\label{eqn:Y_Pop_Mean1} \\
\mu_{y,2} 
& = 
\frac{1}{1-a-b} 
\int_{a}^{1-b}
\left[
h_{Y}
\left( 
F_{Y}^{-1}(s)
\right)\right]^{2} \, ds \nonumber \\
& = 
\frac{1}{1-a-b} 
\int_{a}^{1-b}
\left[F_{X}^{-1} \big( s + (1-s) F_{X}(t \, | \, \bm{\theta})|\bm{\theta}\big) \right]^{2} \, ds \nonumber \\
& = 
\theta^2+2\theta \sigma c_{y, 1} 
+ \sigma^2 c_{y, 2},
\label{eqn:Y_Pop_Mean2}
\end{align}
where for $k=1,2$;
\begin{align}
\label{eq:PPMTMNormalCkdefine}
c_{y,k} & \equiv c_{y, k}(\Phi,a,b,t) 
:= 
\frac{1}{1-a-b}
\int_{a}^{1-b}
\left[\Phi^{-1}\left(s+(1-s)\Phi(\gamma)\right)\right]^{k} \, ds.
\end{align}
Note that $c_{y, k}$ depends on the unknown parameters
but does not depend on the parameters to be estimated for 
completely observed sample data
\citep[see, e.g.,][]{MR2497558}.
Equating 
$\mu_{y,1} 
= 
\widehat{\mu}_{y,1}$ and $\mu_{y,2} 
= 
\widehat{\mu}_{y,2}
$ 
yield the implicit system of equations to be solved for 
$\theta$ and $\sigma$:
\begin{align}
\label{eqn:Normal_MTMPPY_Eqns}
\left\{
\begin{array}{lll}
\theta 
& = &
\widehat{\mu}_{y,1} - c_{y, 1}\sigma 
=: g_{1}(\widehat{\mu}_{y,1},\widehat{\mu}_{y,2}), \\[10pt]
{\sigma} & = & \sqrt{(\widehat{\mu}_{y,2}-\widehat{\mu}_{y,1}^{2})/(c_{y, 2}-c_{y, 1}^{2})} =: g_{2}(\widehat{\mu}_{y,1},\widehat{\mu}_{y,2}).
\end{array}
\right.
\end{align}
The system of Equations (\ref{eqn:Normal_MTMPPY_Eqns}) can be solved for $\widehat{\theta}_{\mbox{\tiny y,MTM}}$ and
$\widehat{\sigma}_{\mbox{\tiny y,MTM}}$ 
using an iterative numerical method with the initializing values:
\begin{align}
\label{eqn:Normal_MTMPPY_EqnsStart}
\sigma_{\mbox{\tiny start}} 
& =
\sqrt{\widehat{\mu}_{y,2} - \widehat{\mu}_{y,1}^{2}} 
\quad \text{and} \quad 
\theta_{\mbox{\tiny start}} 
= 
\widehat{\mu}_{y,1}.
\end{align}

From Theorem \ref{thm:PPY_MTM_AS1},
the entries of the variance-covariance matrix $\bm{\Sigma}_{y}$ 
calculated using (\ref{eq:mtmVB_var_cov}) are
\begin{align*}
\sigma_{11}^{2} 
& = 
\sigma^2
c_{y,1}^{*}, 
\quad 
\sigma_{12}^{2} 
= 
2\theta \sigma^2 c_{y,1}^{*}
+ 
2 \sigma^3c_{y,2}^{*} 
\quad \mbox{and} \quad 
\sigma_{22}^{2} 
= 
4\theta^2 \sigma^2c_{y,1}^{*}
+ 8 \theta \sigma^3c_{y,2}^{*}
+ 4 \sigma^4c_{y,3}^{*},
\end{align*}
where the expressions for $c_{y,k}^{*}, \ k = 1,2,3$
are listed in Appendix \ref{sec:Appendix}.
For $k = 1,2$; it follows evidently that
\begin{align}
\label{eq:PPDerCks}
\left\{
\begin{array}{lll}
\frac{\partial c_{y,k}}{\partial \theta} 
& = &  
-\frac{2^{k-1}\phi(\gamma)}{\sigma (1-a-b)} 	{\displaystyle\int_{a}^{1-b} \frac{(1-s)\left[\Phi^{-1}\left(s+(1-s)\Phi(\gamma)\right)\right]^{k-1}}{\phi\left[\Phi^{-1}\left(s+(1-s)\Phi(\gamma)\right)\right]} \, ds}, \\[15pt]
\frac{\partial c_{y,k}}{\partial \sigma} 
& = &  
-\frac{2^{k-1}(t-\theta)\phi(\gamma)}{\sigma^2 (1-a-b)} {\displaystyle\int_{a}^{1-b} \frac{(1-s)\left[\Phi^{-1}\left(s+(1-s)\Phi(\gamma)\right)\right]^{k-1}}{\phi\left[\Phi^{-1}\left(s+(1-s)\Phi(\gamma)\right)\right]} \, ds}.
\end{array}
\right.
\end{align}
For $k=1,2$; let us denote
\begin{equation*}
\theta_{\mu_{y,k}} 
:=
\left. \frac{\partial g_{1}}{\partial \widehat{\mu}_{y,k}}\right|_{(\mu_{y,1};\mu_{y,2})}
\quad \text{and} \quad
\sigma_{\mu_{y,k}} 
:= 
\left. \frac{\partial g_{2}}{\partial \widehat{\mu}_{y,k}}\right|_{(\mu_{y,1};\mu_{y,2})}.
\end{equation*}
Consider the following more notations:
\begin{align*}
f_{11}(\theta, \sigma) 
& :=
1+\sigma \frac{\partial c_{y,1}}{\partial \theta}, 
& f_{12}(\theta, \sigma) 
& := 
c_{y,1}+\sigma \frac{\partial c_{y,1}}{\partial \sigma}, \\
f_{21}(\theta, \sigma) 
& :=
\frac{\partial c_{y,2}}{\partial \theta}-2c_{y,1}\frac{\partial c_{y,1}}{\partial \theta}, 
& f_{22}(\theta, \sigma) 
& := 
\frac{\partial c_{y,2}}
{\partial \sigma}-2c_{y,1}
\frac{\partial c_{y,1}}{\partial \sigma}.
\end{align*}
The entries of the matrix $\bm{D}_{y}$, given by Theorem \ref{thm:PPY_MTM_AS1},
are found by implicitly differentiating the functions $g_{j}$ (with multivariate chain rule) from Equations 
(\ref{eqn:Normal_MTMPPY_Eqns}) with the help of Equations (\ref{eq:PPDerCks}):
\begin{align*}
d_{11} 
& =
\theta_{\mu_{y,1}} 
= 
\frac{1-f_{12}\sigma_{\mu_{y,1}}}{f_{11}} 
= 
\frac{1-f_{12}d_{21}}{f_{11}}, \\[10pt]
d_{12} 
& = 
\theta_{\mu_{y,2}}
=
-\frac{f_{12}\sigma_{\mu_{y,2}}}{f_{11}} 
=
-\frac{f_{12}d_{22}}{f_{11}}, \\[10pt]
d_{21} 
& =
\sigma_{\mu_{y,1}} 
=
-\frac{K\left[2f_{11}\mu_{y,1}(c_{y,2}-c_{y,1}^{2}) 
+ f_{21}(\mu_{y,2}-\mu_{y,1}^{2})\right]}{f_{11}(c_{y,2}
-c_{y,1}^{2})^{2}+K(\mu_{y,2}-\mu_{y,1}^{2})(f_{11}f_{22}
-f_{12}f_{21})}, \\[10pt]
d_{22} 
& =
\sigma_{\mu_{y,2}} 
=
\frac{Kf_{11}(c_{y,2}-c_{y,1}^{2})}{f_{11}(c_{y,2}
-c_{y,1}^{2})^{2}+K(\mu_{y,2}
-\mu_{y,1}^{2})(f_{11}f_{22}-f_{12}f_{21})}, 
\end{align*}
where 
$K 
:=
\frac{1}{2}
\sqrt{\frac{c_{y,2}-c_{y,1}^{2}}{\mu_{y,2}-\mu_{y,1}^{2}}}$.
Hence, the asymptotic result (\ref{eqn:MTM_PPY_AsymResult1})
becomes
\begin{equation}
\label{eq:PP_MTMEstNormal}
(\widehat{\theta}_{\mbox{\tiny y,MTM}},
\widehat{\sigma}_{\mbox{\tiny y,MTM}})
\sim \mathcal{AN}
\left((\theta,\sigma),n^{-1}\bm{S}_{\mbox{\tiny y,MTM}}\right), \ 
\bm{S}_{\mbox{\tiny y,MTM}}
:= 
\bm{D}_{y}\bm{\Sigma}_{y} \bm{D}_{y}^{'}.
\end{equation}
From (\ref{eqn:MLEYDeltaMatrix2}) and 
(\ref{eq:PP_MTMEstNormal}), 
it follows that
\begin{align}
\label{eqn:PPY_MTM_MLE_ARE}
\mbox{ARE}
\left( 
\left(\widehat{\theta}_{\mbox{\tiny y,MTM}},
\widehat{\sigma}_{\mbox{\tiny y,MTM}}\right),
\left(\widehat{\theta}_{\mbox{\tiny y,MLE}},\widehat{\sigma}_{\mbox{\tiny y,MLE}}\right)
\right)
& =
\left(
{\mbox{det}\left(\bm{S}_{\mbox{\tiny y,MLE}}\right)}/
{\mbox{det}
\left(
\bm{S}_{\mbox{\tiny y,MTM}}
\right)}
\right)^{0.5}.
\end{align}
For some selected trimming proportions $a$ and $b$,
numerical values of the AREs given by 
Equation \eqref{eqn:PPY_MTM_MLE_ARE}
are provided in Table \ref{table:PPY_MLE_MTM_ARE}.

\begin{table}[tbh!]
\centering
\caption{
$
\mbox{ARE}
\left( 
\left(\widehat{\theta}_{\mbox{\tiny y,MTM}},
\widehat{\sigma}_{\mbox{\tiny y,MTM}}\right),
\left(\widehat{\theta}_{\mbox{\tiny y,MLE}},
\widehat{\sigma}_{\mbox{\tiny y,MLE}}\right)
\right)
$ 
for fixed $d = 4$ and selected $a$ and $b$ and various choices of
right-censoring point $u$ from $\mbox{LN}(1,5,3)$.
}
\label{table:PPY_MLE_MTM_ARE}
\begin{tabular}{c|ccccc|cccc|ccc}
\hline
\multicolumn{1}{c|}{} &
\multicolumn{5}{|c|}{$b$ (when $u=2\times10^{5}$)} &
\multicolumn{4}{|c|}{$b$ (when $u=2.4\times10^{4}$)} &
\multicolumn{3}{|c}{$b$ (when $u=8.5\times10^{3}$)} \\[-0.5ex]
\cline{2-13}
$a$ & 0.01 & 0.05 & 0.10 & 0.15 & 0.25 & 
0.05 & 0.10 & 0.15 & 0.25 & 0.10 & 0.15 & 0.25 \\
\hline
\hline
0 & 0.987 & 0.904 & 0.821 & 0.747 & 0.616 & 
0.960 & 0.871 & 0.793 & 0.654 & 0.934 & 0.850 & 0.701 \\
0.05 & 0.984 & 0.904 & 0.821 & 0.749 & 0.620 & 
0.959 & 0.872 & 0.795 & 0.658 & 0.935 & 0.852 & 0.705 \\
0.10 & 0.971 & 0.893 & 0.813 & 0.742 & 0.615 & 
0.948 & 0.863 & 0.788 & 0.653 & 0.925 & 0.844 & 0.700 \\
0.15 & 0.948 & 0.874 & 0.796 & 0.726 & 0.602 & 
0.927 & 0.845 & 0.771 & 0.639 & 0.906 & 0.827 & 0.685 \\
0.25 & 0.885 & 0.816 & 0.742 & 0.676 & 0.556 & 
0.867 & 0.788 & 0.718 & 0.590 & 0.845 & 0.769 & 0.633 \\
\hline
\end{tabular}
\end{table}

\begin{note}
\label{note:PPY_MTM_Symplified1}
If we empirically estimate $\Phi(\gamma)$ 
in Equation (\ref{eq:PPMTMNormalCkdefine})
using $\sigma_{\mbox{\tiny start}}$ 
and $\theta_{\mbox{\tiny start}}$ 
given by (\ref{eqn:Normal_MTMPPY_EqnsStart}),
then $c_{y,k}$ given by (\ref{eq:PPMTMNormalCkdefine}) 
is no longer a function of 
the parameters $\theta$ and $\sigma$ to 
be estimated and hence the explicit estimated
values of $\theta$ and $\sigma$ are given 
by the system (\ref{eqn:Normal_MTMPPY_Eqns})
which is equivalent to the complete data scenario.
This technique could be implemented to 
eliminate all the computational complexity 
we discussed in this section.
\qed 
\end{note}

\subsection{Payments {\em Z}}
\label{sec:PPZ_MTM1}

From (\ref{p2NZdata}), it follows that payment $Z$ is 
left- and right-censored form of complete  
random variable $X$.
Thus, possible permutations between 
$a$, $b$, and their positioning with respect to 
$F_{X}(t)$ and $F_{X}(T)$ have to be taken into account,
since the expressions for $\sigma_{ij}^{2}$ given by 
(\ref{eq:mtmVB_var_cov}) with qf (\ref{p2qf}) depend
on the six possible permutations among $a$, $b$,
$F_{X}(t)$, and $F_{X}(T)$:
\begin{enumerate}
\item $0 \leq a < 1-b \leq F_{X}(t) < F_{X}(T) \leq 1$. \qquad \qquad 
4. $0 \leq F_{X}(t) < F_{X}(T) \leq a < 1-b \leq 1$.
\item $0 \leq a \leq F_{X}(t) < 1-b \leq F_{X}(T) \leq 1$. \qquad \qquad 
5. $0 \leq F_{X}(t) \leq a < F_{X}(T) \leq 1-b \leq 1$.
\item $0 \leq a \leq F_{X}(t) < F_{X}(T) \leq 1-b \leq 1$. \qquad \qquad 
6. $0 \leq F_{X}(t) \leq a < 1-b \leq F_{X}(T) \leq 1$.
\end{enumerate}
Among these six cases, two of those scenarios 
(estimation based on censored data only -- Cases 1 and 4)
have no parameters to be estimated in the formulas of population 
trimmed moments and three 
(estimation based on observed and censored data -- Cases 2, 3, and 5)
are inferior to the estimation scenario based on fully observed data. 
Thus, from now on, we will proceed only with Case 6 which makes
most sense and simplifies the estimation procedure significantly 
because it uses the available data in the most effective way. 
Moreover, the MTM estimators based on Case 6 will be resistant 
to outliers, that is, observations that are inconsistent with the 
assumed model and most likely appearing at the boundaries $t$ and $T$.
Case 6 also eliminates heavier point masses given at 
the censored points $t$ and $T$. 

\bigskip

\begin{note}
The MTM estimators with $a>0$
and 
$
b>0
$ 
$
(0 \leq F_{X}(t_1) \leq a < 1-b \leq F_{X}(t_2) \leq 1)
$ 
are globally 
robust with the {\em lower\/} and {\em upper\/} breakdown points given by
$\mbox{\sc lbp} = a$ and $\mbox{\sc ubp} = b$, respectively. The robustness
of such estimators against small or large outliers comes from the fact that
in the computation of estimates the influence of the order statistics with
the index less than $n \times \mbox{\sc lbp}$ or higher than 
$n \times (1- \mbox{\sc ubp})$ is limited. For more details on {\sc lbp} 
and {\sc ubp}, see \cite{MR1989836} and \cite{MR1987777}.
\qed
\end{note}

For practical data analysis purpose, standard empirical estimates of 
$F_{X}(t)$ and $F_{X}(T)$ provide guidance about the choice of $a$ and 
$1-b$ and are chosen according to
\begin{equation}
\label{eqn:PPZ_Cond1}
F_{n}(t) 
\le a 
< 1-b 
\le F_{n}(T), 
\quad \mbox{where} \quad F_{n} \ 
\mbox{is the empirical cdf.}
\end{equation}
Further, define 
$h_Z(z) := z/c+t$. 
Now, consider an observed {\em i.i.d.\/} sample
$z_{1},\ldots,z_{n}$ defined by 
(\ref{p2qf}).
Let $z_{1:n}, \ldots, z_{n:n}$ be the corresponding 
ordered statistics. 
Then, the sample trimmed moments given by (\ref{Ts})
are given by
\begin{align}
\label{eqn:PPZMTMHat1}
\left\{
\begin{array}{lcl}
\widehat{\mu}_{z,1} 
& = &
\displaystyle 
\frac{1}{n-m_{n}-m_{n}^{*}}
\sum_{i=m_{n}+1}^{n-m_{n}^{*}}
h_{Z} (z_{i:n})
=
\frac{1}{n-m_{n}-m_{n}^{*}}
\sum_{i=m_{n}+1}^{n-m_{n}^{*}}
\left(\frac{z_{i:n}}{c}+t\right), \\[20pt]
\widehat{\mu}_{z,2}
& = &
\displaystyle  
\frac{1}{n-m_{n}-m_{n}^{*}}
\sum_{i=m_{n}+1}^{n-m_{n}^{*}}
\left(h_{Z} (z_{i:n})\right)^{2}
=
\frac{1}{n-m_{n}-m_{n}^{*}}
\sum_{i=m_{n}+1}^{n-m_{n}^{*}}{\left(\frac{z_{i:n}}{c}+t\right)^{2}},
\end{array}
\right.
\end{align}
with $m_{n}/n \rightarrow a$ and $m_{n}^{*}/n \rightarrow b$. 
With Case 6, choose
$m_{n} \geq \sum_{i=1}^{n}\ID\{z_{i} = 0\}$ and 
$m_{n}^{*} \geq \sum_{i=1}^{n}\ID\{z_{i} = cR\}$.
By assuming the most general case that 
$0 \leq F_{X}(t) \leq a < 1-b \leq F_{X}(T) \leq 1$,
the corresponding population trimmed moments (\ref{Tp})
with the qf (\ref{p2qf}) are given by:
\begin{align*}
\mu_{z,1} 
& :=
\frac{1}{1-a-b}\int_{a}^{1-b}{h_{Z}\left(F_{Z}^{-1}(s)\right)} \, ds \\
& = 
\frac{1}{1-a-b}\int_{a}^{1-b}{F_{X}^{-1}(s \, | \, \bm{\theta})} \, ds \\
& = 
\theta +\sigma c_{1}, \\
\mu_{z,2} 
& :=
\frac{1}{1-a-b}\int_{a}^{1-b}
\left[h_{Z}\left({F_{Z}^{-1}(s)}\right)\right]^{2} \, ds \\
& = \frac{1}{1-a-b}\int_{a}^{1-b}
\left[{F_{X}^{-1}(s \, | \,
\bm{\theta})}\right]^{2} \, ds \\
& = \theta^2 +2\theta\sigma c_{1} +\sigma^2 c_{2},
\end{align*}
where 
$c_{k} \equiv c_{y,k}, \ 1 \le k \le 4$,
given by (\ref{eq:PPMTMNormalCkdefine})
with $\gamma=-\infty$ 
and are listed in Appendix \ref{sec:Appendix}.
Thus, with the assumption 
$0 \leq F_{X}(t) \leq a < 1-b \leq F_{X}(T) \leq 1$, 
this case translates to the complete data
case which is fully investigated by \cite{MR2497558}
and the the MTM estimators of $\theta$ and $\sigma$ are
\begin{align}
\label{eqn:PPY_MTM_Estimators1}
\left\{
\begin{array}{rcl}
\widehat{\theta}_{\mbox{\tiny z,MTM}} 
& = & 
\widehat{\mu}_{z,1}
-c_{1} \, \widehat{\sigma}_{\mbox{\tiny z,MTM}} \\[5pt]
\widehat{\sigma}_{\mbox{\tiny z,MTM}}
& = & 
\sqrt{
	\left(\widehat{\mu}_{z,2} - \widehat{\mu}_{z,1}^{2}\right)/
	\left(c_{2}-c_{1}^{2}\right)}
\end{array}
\right.
\end{align}
And, the corresponding ARE is given by:
\begin{align}
\label{eqn:PPZ_MTM_MLE_ARE}
\mbox{ARE}
\left( 
\left(\widehat{\theta}_{\mbox{\tiny z,MTM}},
\widehat{\sigma}_{\mbox{\tiny z,MTM}}\right),
\left(\widehat{\theta}_{\mbox{\tiny z,MLE}},
\widehat{\sigma}_{\mbox{\tiny z,MLE}}\right)
\right)
& =
\left(
{\mbox{det}\left(\bm{S}_{\mbox{\tiny z,MLE}}\right)}/
{\mbox{det}
\left(
\bm{S}_{\mbox{\tiny z,MTM}}
\right)}
\right)^{0.5},
\end{align}
where 
\[
\bm{S}_{\mbox{\tiny z,MTM}}
:= 
\dfrac{\sigma^2}{\left(c_{2}-c_{1}^{2}\right)^{2}}
\begin{bmatrix}
c_{1}^{*}c_{2}^{2}-2c_{1}c_{2}c_{2}^{*}+c_{1}^{2}c_{3}^{*} & 
-c_{1}^{*}c_{1}c_{2}+c_{2}c_{2}^{*}+c_{1}^{2}c_{2}^{*}-c_{1}c_{3}^{*} \\[5pt]
-c_{1}^{*}c_{1}c_{2}+c_{2}c_{2}^{*}+c_{1}^{2}c_{2}^{*}-c_{1}c_{3}^{*} & 
c_{1}^{*}c_{1}^{2}-2c_{1}c_{2}^{*}+c_{3}^{*} 
\end{bmatrix},
\]
and the expressions for 
$c_{k}^{*}, \ k = 1,2,3$, as functions of 
$a,b,c_{1},c_{2},c_{3}$, and $c_{4}$ 
are such that
$
c_{k}^{*}
\equiv 
c_{y,k}^{*}
$
with $\gamma = -\infty$
and are listed in Appendix \ref{sec:Appendix}.
Numerical values of the AREs given by 
Equation \eqref{eqn:PPZ_MTM_MLE_ARE}
are summarized in 
Table \ref{table:PPZ_MLE_MTM_ARE}
for some selected values of $a$ and $b$. 

\begin{table}[tbh!]
\centering
\caption{
$
\mbox{ARE}
\left( 
\left(\widehat{\theta}_{\mbox{\tiny z,MTM}},
\widehat{\sigma}_{\mbox{\tiny z,MTM}}\right),
\left(\widehat{\theta}_{\mbox{\tiny z,MLE}},
\widehat{\sigma}_{\mbox{\tiny z,MLE}}\right)
\right)
$ 
for fixed $d = 4$ and selected $a$ and $b$ and various choices of
right-censoring point $u$ from $\mbox{LN}(1,5,3)$.
}
\label{table:PPZ_MLE_MTM_ARE}
\begin{tabular}{c|ccccc|cccc|ccc}
\hline
\multicolumn{1}{c|}{} &
\multicolumn{5}{|c|}{$b$ (when $u=2\times10^{5}$)} &
\multicolumn{4}{|c|}{$b$ (when $u=2.4\times10^{4}$)} &
\multicolumn{3}{|c}{$b$ (when $u=8.5\times10^{3}$)} \\[-0.5ex]
\cline{2-13}
$a$ & 0.01 & 0.05 & 0.10 & 0.15 & 0.25 & 
0.05 & 0.10 & 0.15 & 0.25 & 0.10 & 0.15 & 0.25 \\
\hline
\hline
0.10 & 0.948 & 0.900 & 0.844 & 0.793 & 0.695 & 
0.933 & 0.876 & 0.822 & 0.720 & 0.914 & 0.858 & 0.752 \\
0.15 & 0.891 & 0.846 & 0.793 & 0.742 & 0.647 & 
0.877 & 0.822 & 0.770 & 0.671 & 0.858 & 0.804 & 0.701 \\
0.25 & 0.786 & 0.745 & 0.695 & 0.647 & 0.556 & 
0.772 & 0.720 & 0.671 & 0.577 & 0.752 & 0.701 & 0.602 \\
0.49 & 0.550 & 0.516 & 0.471 & 0.428 & 0.343 & 
0.535 & 0.489 & 0.444 & 0.355 & 0.510 & 0.464 & 0.371 \\
\hline
\end{tabular}
\end{table}

\section{Special Cases}
\label{sec:SpecialCases}

Singly left truncated, that is, 
$u \to \infty$,
which is equivalent to 
$T \to \infty$,
sample data set is very common in insurance industries as well
as in operational risk modeling \citep[see, e.g.,][]{EPUS}.
So, in this section, we summarize the analogous results
from Sections \ref{sec:MLE} and \ref{sec:MTM} for 
singly left truncated lognormal sample data. 
Clearly, for $T \to \infty$ implies
$
\Omega_{2} = 0, \
\Omega_{y,2} = 0, \
\mbox{and} \
\Omega_{z,2} = 0.
$
Thus, the system of Equations (\ref{eqn:PPNormalMLEeqn2})
becomes
\begin{align}
\label{eqn:PPNormalMLEeqnLT1}
\left\{
\begin{array}{rrr}
\sigma\left(\Omega_{y,1}-\gamma\right) 
- c^{-1}\widehat{\mu}_{y,1} 
& = & 0, \\[10pt]
\sigma^{2}\left(1-\gamma(\Omega_{y,1}-\gamma)\right)
-  c^{-2}\widehat{\mu}_{y,2} 
& = & 0,
\end{array} \right.
\end{align}
The system (\ref{eqn:PPNormalMLEeqnLT1}) can be rearranged
as a nonlinear equation of $\gamma$ only as:
\begin{align}
\label{eqn:PPNormalMLEeqnLT2}
G(\gamma) 
& = 
\delta_{y}
\quad \mbox{where} \quad 
G(\gamma) 
:= 
\dfrac{1}{\Omega_{y,1}-\gamma} 
\left[ 
\dfrac{1}{\Omega_{y,1}-\gamma} - \gamma 
\right], \ 
\Omega_{y,1} 
=
\dfrac{\phi(\gamma)}{\bar{\Phi}(\gamma)}, \
\ \mbox{and} \
\delta_{y}
:=
\dfrac{\widehat{\mu}_{y,2}}{\widehat{\mu}_{y,1}^{2}}
\end{align}
It is important to note that equation 
(\ref{eqn:PPNormalMLEeqnLT2}) is a nonlinear equation 
to be solved for $\gamma$, but the system 
(\ref{eqn:PPNormalMLEeqn2}) should be solved simultaneously 
both for $\gamma$ and $\sigma$.
It was shown by \cite{MR61319} that the function $G(\cdot)$
is monotonically increasing on the real line 
$\mathbb{R}$ with 
$
\displaystyle 
\lim_{\gamma \to -\infty}G(\gamma) = 1
$
and 
$
\displaystyle 
\lim_{\gamma \to \infty}G(\gamma) = 2.
$
Thus, from Equation (\ref{eqn:PPNormalMLEeqnLT2}),
we have:

\begin{thm}
\label{thm:leftTruncatedUniqueSol1}
The equation $G(\gamma) = \delta_{y}$
has a unique solution if and only if 
$1 < \delta_{y} < 2$ and the corresponding unique solution 
$(\widehat{\gamma}_{\mbox{\tiny y,MLE}},
\widehat{\sigma}_{\mbox{\tiny y,MLE}})$ 
of the system (\ref{eqn:PPNormalMLEeqnLT1})
is given by:
\begin{align}
\label{eqn:leftTruncatedUniqueSol1}
(\widehat{\gamma}_{\mbox{\tiny y,MLE}},
\widehat{\sigma}_{\mbox{\tiny y,MLE}})
& = 
\left( 
G^{-1}(\delta_{y}),
\dfrac{c^{-1}\widehat{\mu}_{y,1}}
{\frac{\phi(\widehat{\gamma}_{\mbox{\tiny y,MLE}})}
	{\bar{\Phi}(\widehat{\gamma}_{\mbox{\tiny y,MLE}})}
	-\widehat{\gamma}_{\mbox{\tiny y,MLE}}}
\right),
\end{align}
and consequently
$
\widehat{\theta}_{\mbox{\tiny y,MLE}} 
= 
t - \widehat{\sigma}_{\mbox{\tiny y,MLE}} \,
\widehat{\gamma}_{\mbox{\tiny y,MLE}}.
$
\end{thm}

Further, it has been established by \cite{EPUS}
that the unique solution 
$(\widehat{\gamma}_{\mbox{\tiny y,MLE}},
\widehat{\sigma}_{\mbox{\tiny y,MLE}})$
given by Theorem \ref{thm:leftTruncatedUniqueSol1}
is, in fact, the point of global maximum
of the log-likelihood surface given by 
equation (\ref{eqn:PPYMLELogLik1}) with the
adjustment of $T \to \infty$.
Therefore, it follows that 
$1 < \delta_{y} < 2$ is the necessary and sufficient 
condition for the system (\ref{eqn:PPNormalMLEeqnLT1})
to have the unique global MLE solution
$(\widehat{\gamma}_{\mbox{\tiny y,MLE}},
\widehat{\sigma}_{\mbox{\tiny y,MLE}})$.

The asymptotic result 
(\ref{eqn:MLEYDeltaMatrix1}) takes the form:
\begin{equation}
\label{eqn:MLEYDeltaMatrix3}
(\widehat{\gamma}_{\mbox{\tiny y,MLE}},
\widehat{\sigma}_{\mbox{\tiny y,MLE}}) \sim \mathcal{AN}\left((\gamma,\sigma),
\frac{1}{n\left(r_{1}r_{3}-r_{2}^{2}\right)}
\begin{bmatrix}
-r_{3} & \sigma r_{2} \\[10pt]
\sigma r_{2} & -\sigma^{2}r_{1} 
\end{bmatrix}
\right),
\end{equation}
where 
\begin{align}
\label{eqn:PPNormalMLEFishergFunctions2}
r_{1}(\gamma,\xi) 
& = 
-\left[1+\gamma\Omega_{1}
-
\frac{\phi(\gamma)}{\bar{\Phi}(\gamma)}\Omega_{1} \right], \quad 
r_{2}(\gamma,\xi) 
= 
\Omega_{1}-\gamma, \quad
r_{3}(\gamma,\xi) 
=
-\left[2 -\gamma(\Omega_{1}-\gamma) \right]. 
\end{align}

Similarly, for payment-per-loss data scenario,
the system (\ref{eqn:censoredNormalMLEeqn2})
takes the form:
\begin{align}
\label{eqn:censoredNormalMLEeqnLT3}
\left\{
\begin{array}{rrr}
\sigma\left(\Omega_{z,1}-\gamma\right) 
- c^{-1}\widehat{\mu}_{z,1} 
& = & 0, \\[10pt]
\sigma^{2}\left(1-\gamma(\Omega_{z,1}-\gamma)\right)
- c^{-2}\widehat{\mu}_{z,2} & = & 0,
\end{array} \right.				
\end{align}
and a nonlinear equation to be solved for
$\gamma$ is given by:
\begin{align}
\label{eqn:PPNormalMLEeqnLTZ}
\dfrac{1}{\Omega_{z,1}-\gamma} 
\left[ 
\dfrac{1}{\Omega_{z,1}-\gamma} - \gamma 
\right]
& = 
\dfrac{\widehat{\mu}_{z,2}}{\widehat{\mu}_{z,1}^{2}}, 
\quad \mbox{where} \quad 
\Omega_{z,1}
=
\dfrac{n_{0}}{n_{1}} \ 
\dfrac{\phi(\gamma)}{\Phi(\gamma)}.
\end{align}
Clearly, the empirical estimates of 
$\Phi(\gamma)$ and $\bar{\Phi}(\gamma)$
are $n_{0}/n$ and $n_{1}/n$, respectively. 
Thus, if we replace the ratios 
$n_{0}/n$ and $n_{1}/n$ by 
$\Phi(\gamma)$ and $\bar{\Phi}(\gamma)$,
respectively, the nonlinear equation 
(\ref{eqn:PPNormalMLEeqnLTZ}) takes the form: 
\begin{align}
\label{eqn:PPNormalMLEeqnPPZ2}
G(\gamma) 
& = 
\delta_{z},
\quad \mbox{where} \quad 
\delta_{z}
:=
\dfrac{\widehat{\mu}_{z,2}}{\widehat{\mu}_{z,1}^{2}}.
\end{align}
Then, with the condition 
$1 < \delta_{z} < 2$, 
Theorem \ref{thm:leftTruncatedUniqueSol1}
is applicable again to get the unique global MLE solution 
of the system (\ref{eqn:censoredNormalMLEeqnLT3})
with equation (\ref{eqn:PPNormalMLEeqnPPZ2}) as:
\begin{align}
\label{eqn:leftTruncatedUniqueSol2}
(\widehat{\gamma}_{\mbox{\tiny z,MLE}},
\widehat{\sigma}_{\mbox{\tiny z,MLE}})
& = 
\left( 
G^{-1}(\delta_{z}),
\dfrac{c^{-1}\widehat{\mu}_{z,1}}
{\frac{\phi(\widehat{\gamma}_{\mbox{\tiny z,MLE}})}
{\bar{\Phi}(\widehat{\gamma}_{\mbox{\tiny z,MLE}})}
-\widehat{\gamma}_{\mbox{\tiny z,MLE}}}
\right),
\end{align}
and consequently
$
\widehat{\theta}_{\mbox{\tiny z,MLE}} 
= 
t - \widehat{\sigma}_{\mbox{\tiny z,MLE}} \,
\widehat{\gamma}_{\mbox{\tiny z,MLE}}$.

Finally, the asymptotic result 
(\ref{eqn:MLEZDeltaMatrix1}) takes the form:
\begin{equation}
\label{eqn:MLEZDeltaMatrix3}
(\widehat{\gamma}_{\mbox{\tiny z,MLE}},
\widehat{\sigma}_{\mbox{\tiny z,MLE}}) 
\sim \mathcal{AN}\left((\gamma,\sigma),
\frac{1}{n \bar{\Phi}(\gamma)
\left(\psi_{1}\psi_{3}-\psi_{2}^{2}\right)}
\begin{bmatrix}
-\psi_{3} & \sigma \psi_{2} \\[10pt]
\sigma \psi_{2} & -\sigma^{2}\psi_{1} 
\end{bmatrix}
\right),
\end{equation}
where 
\begin{align}
\label{eqn:censoredNormalMLEFishergFunctions2}
\psi_{1}(\gamma,\xi) 
=
-\left[1+\gamma\Omega_{1}
+\frac{\phi(\gamma)}{\Phi(\gamma)}\Omega_{1}  \right], \quad 
\psi_{2}(\gamma,\xi) 
=
\Omega_{1}-\gamma, \quad
\psi_{3}(\gamma,\xi)
=
-\left[2 -\gamma(\Omega_{1}-\gamma)\right].
\end{align}

Now, we discuss the analogous results for MTM approach 
for singly left truncated lognormal data scenarios.
For payments $Y$, since $T \to \infty$, then the value of
$s^{*}$ as defined in Section \ref{sec:PPY_MTM1} is simply 
1 which is Case 2 as discussed in Section \ref{sec:PPY_MTM1} 
and the entire theory developed in Section \ref{sec:PPY_MTM1} 
is for Case 2. 

Similarly, for payments $Z$ with $T \to \infty$,
all the six possible permutations mentioned
in Section \ref{sec:PPZ_MTM1} shrank to only
three Cases 1, 2, and 6. 
Further, with a goal of finding the most resistant
estimators we discussed only Case 6 in Section 
\ref{sec:PPZ_MTM1} and the same results also
hold for $T \to \infty$.

\section{Simulation Study} 
\label{sec:SimStudy}

This section supplements the theoretical results 
we developed in Section \ref{sec:MTM} via simulation. 
The main goal is to access the size of the sample such that the
estimators are free from bias 
(given that the estimators are asymptotically unbiased),
justify the asymptotic normality, and
their finite sample relative efficiencies (REs)  
to reach the corresponding AREs. 
To compute RE of MTM estimators 
we use MLE as a benchmark. 
Thus, the definition of ARE given by Equation 
(\ref{eq:infinite_relative_efficiency_benchmark_MLE})
for finite sample performance translates to:
\begin{equation} 
\label{eq:finite_relative_efficiency_benchmark_MLE}
RE(\mbox{MTM, MLE}) 
=
\frac{\text{asymptotic variance of MLE estimator}}
{\text{small-sample variance of a competing MTM estimator}},
\end{equation}
where the numerator is as defined in 
(\ref{eq:infinite_relative_efficiency_benchmark_MLE})
and the denominator is given by:
\[
\left(
\mbox{det}
\begin{bmatrix}
E 
\left[ 
\left(\widehat{\theta} - \theta \right)^{2}
\right] & 
E 
\left[ 
\left(\widehat{\theta} - \theta \right)
\left(\widehat{\sigma} - \sigma \right)
\right] \\[15pt]
E 
\left[ 
\left(\widehat{\theta} - \theta \right)
\left(\widehat{\sigma} - \sigma \right)
\right] & 
E 
\left[ 
\left(\widehat{\sigma} - \sigma \right)^{2}
\right] 
\end{bmatrix}
\right)^{1/2}.
\]

\begin{table}[hbt!]
\caption{
Lognormal payment-per-payment 
actuarial loss scenario,
$LN(w_{0}=1,\theta=5,\sigma=3)$ with $d = 4$
and two selected values of right-censoring point $u$.
The entries are mean values 
based on 10,000 samples.
}
\label{table:PPY_SimStudy2}
\centering
\begin{tabular}{c|cc|cc|cc|cc|cc|cc}
\cline{2-13}
{} & 
\multicolumn{2}{c|}{Proportion} & 
\multicolumn{2}{|c|}{$n=100$} & 
\multicolumn{2}{|c|}{$n=250$} & 
\multicolumn{2}{|c|}{$n=500$} &
\multicolumn{2}{|c|}{$n=1000$} &
\multicolumn{2}{|c}{$n \to \infty$} \\
\cline{2-13} 
& & & & & & & & & & & \\[-2.50ex]
{} & 
$a$ & $b$ & 
$\widehat{\theta}/\theta$ & 
$\widehat{\sigma}/\sigma$ &
$\widehat{\theta}/\theta$ & 
$\widehat{\sigma}/\sigma$ &
$\widehat{\theta}/\theta$ & 
$\widehat{\sigma}/\sigma$ &
$\widehat{\theta}/\theta$ & 
$\widehat{\sigma}/\sigma$ &
$\widehat{\theta}/\theta$ & 
$\widehat{\sigma}/\sigma$ \\
\hline\hline 
\multicolumn{1}{c|}{} &
\multicolumn{12}{c}{} \\[-2.50ex]
\multirow{16}{*}{\rotatebox{90}{$u=2 \times 10^{5}$}} &
\multicolumn{12}{c}
{Mean values of
	$\widehat{\theta}/\theta$ 
	and
	$\widehat{\sigma}/\sigma$.} \\
\cline{2-13} 
{} & 
\multicolumn{2}{|c|}{MLE} &
0.99 & 1.00 & 
1.00 & 1.00 & 
1.00 & 1.00 & 
1.00 & 1.00 & 
1 & 1 \\
{} & 
0.05 & 0.05 &
0.99 & 1.01 & 
0.99 & 1.02 &
1.00 & 1.00 &
1.00 & 1.00 &
1 & 1 \\
{} & 
0.10 & 0.10 &
0.98 & 1.02 & 
0.99 & 1.01 &
1.00 & 1.00 &
1.00 & 1.00 &
1 & 1 \\
{} & 
0.15 & 0.15 &
0.98 & 1.02 & 
0.99 & 1.02 &
1.00 & 1.00 &
1.00 & 1.00 &
1 & 1 \\
\cline{2-13}
{} & 
0.00 & 0.05 &
0.99 & 1.01 & 
1.00 & 1.00 &
1.00 & 1.00 &
1.00 & 1.00 &
1 & 1 \\
{} &
0.00 & 0.10 & 
0.99 & 1.01 & 
0.99 & 1.00 &
1.00 & 1.00 & 
1.00 & 1.00 & 
1 & 1 \\
{} & 
0.00 & 0.25 &
0.98 & 1.02 & 
0.99 & 1.01 &
1.00 & 1.00 &
1.00 & 1.00 &
1 & 1 \\
\cline{2-13}
\multicolumn{1}{c|}{} &
\multicolumn{10}{c}{} \\[-2.50ex]
\multicolumn{1}{c|}{} &
\multicolumn{12}{c}
{Finite-sample efficiencies of MTMs relative to MLEs.} \\
\cline{2-13}
{} & 
\multicolumn{2}{|c|}{MLE} &
\multicolumn{2}{|c}{0.93} & 
\multicolumn{2}{|c}{0.99} &
\multicolumn{2}{|c}{0.98} &
\multicolumn{2}{|c}{1.00} &
\multicolumn{2}{|c}{1} \\
{} & 
0.05 & 0.05 &
\multicolumn{2}{|c|}{0.80} & 
\multicolumn{2}{|c|}{0.85} &
\multicolumn{2}{|c|}{0.89} &
\multicolumn{2}{|c|}{0.90} &
\multicolumn{2}{|c}{0.904} \\
{} & 
0.10 & 0.10 &
\multicolumn{2}{|c|}{0.69} & 
\multicolumn{2}{|c|}{0.78} &
\multicolumn{2}{|c|}{0.79} &
\multicolumn{2}{|c|}{0.80} &
\multicolumn{2}{|c}{0.813} \\
{} & 
0.15 & 0.15 &
\multicolumn{2}{|c|}{0.57} & 
\multicolumn{2}{|c|}{0.66} &
\multicolumn{2}{|c|}{0.70} &
\multicolumn{2}{|c|}{0.71} &
\multicolumn{2}{|c}{0.726} \\
\cline{2-13}
{} & 
0.00 & 0.05 &
\multicolumn{2}{|c|}{0.80} & 
\multicolumn{2}{|c|}{0.87} &
\multicolumn{2}{|c|}{0.89} &
\multicolumn{2}{|c|}{0.90} & 
\multicolumn{2}{|c}{0.904} \\
{} &
0.00 & 0.10 & 
\multicolumn{2}{|c|}{0.70} & 
\multicolumn{2}{|c|}{0.79} &
\multicolumn{2}{|c|}{0.81} & 
\multicolumn{2}{|c|}{0.81} &
\multicolumn{2}{|c}{0.821} \\
{} & 
0.00 & 0.25 &
\multicolumn{2}{|c}{0.41} & 
\multicolumn{2}{|c|}{0.56} &
\multicolumn{2}{|c|}{0.59} &
\multicolumn{2}{|c|}{0.60} &
\multicolumn{2}{|c}{0.616} \\
\hline\hline 
\multicolumn{1}{c|}{} &
\multicolumn{10}{c}{} \\[-2.50ex]
\multirow{14}{*}{\rotatebox{90}{$u=2.4 \times 10^{4}$}} &
\multicolumn{12}{c}
{Mean values of
	$\widehat{\theta}/\theta$ 
	and
	$\widehat{\sigma}/\sigma$.} \\
\cline{2-13} 
{} & 
\multicolumn{2}{|c|}{MLE} &
0.99 & 1.00 & 
1.00 & 1.00 & 
1.00 & 1.00 & 
1.00 & 1.00 & 
1 & 1 \\
{} & 
0.10 & 0.10 &
0.98 & 1.02 & 
0.99 & 1.01 &
1.00 & 1.00 &
1.00 & 1.00 & 
1 & 1 \\
{} & 
0.15 & 0.15 &
0.98 & 1.02 & 
0.99 & 1.02 &
1.00 & 1.00 &
1.00 & 1.00 & 
1 & 1 \\
\cline{2-13}
{} & 
0.00 & 0.10 &
0.99 & 1.01 & 
0.99 & 1.00 &
1.00 & 1.00 &
1.00 & 1.00 & 
1 & 1 \\
{} & 
0.05 & 0.10 &
0.98 & 1.01 & 
0.99 & 1.01 &
1.00 & 1.00 &
1.00 & 1.00 & 
1 & 1 \\
{} & 
0.00 & 0.25 &
0.98 & 1.02 & 
0.99 & 1.01 &
1.00 & 1.00 &
1.00 & 1.00 & 
1 & 1 \\
\cline{2-13}
\multicolumn{1}{c|}{} &
\multicolumn{10}{c}{} \\[-2.50ex]
\multicolumn{1}{c|}{} &
\multicolumn{12}{c}
{Finite-sample efficiencies of MTMs relative to MLEs.} \\
\cline{2-13}
{} & 
\multicolumn{2}{|c|}{MLE} &
\multicolumn{2}{|c|}{0.93} & 
\multicolumn{2}{|c|}{0.96} &
\multicolumn{2}{|c|}{0.98} &
\multicolumn{2}{|c|}{1.00} &
\multicolumn{2}{|c}{1} \\
{} & 
0.10 & 0.10 &
\multicolumn{2}{|c|}{0.76} &
\multicolumn{2}{|c|}{0.81} &
\multicolumn{2}{|c|}{0.84} &
\multicolumn{2}{|c|}{0.86} &
\multicolumn{2}{|c}{0.863} \\
{} & 
0.15 & 0.15 &
\multicolumn{2}{|c|}{0.65} &
\multicolumn{2}{|c|}{0.69} &
\multicolumn{2}{|c|}{0.74} &
\multicolumn{2}{|c|}{0.77} &
\multicolumn{2}{|c}{0.771} \\
\cline{2-13}
{} & 
0.00 & 0.10 &
\multicolumn{2}{|c|}{0.77} &
\multicolumn{2}{|c|}{0.83} &
\multicolumn{2}{|c|}{0.85} &
\multicolumn{2}{|c|}{0.87} &
\multicolumn{2}{|c}{0.871} \\
{} & 
0.05 & 0.10 &
\multicolumn{2}{|c|}{0.77} & 
\multicolumn{2}{|c|}{0.80} &
\multicolumn{2}{|c|}{0.85} &
\multicolumn{2}{|c|}{0.87} &
\multicolumn{2}{|c}{0.872} \\
{} & 
0.00 & 0.25 &
\multicolumn{2}{|c|}{0.51} & 
\multicolumn{2}{|c|}{0.61} &
\multicolumn{2}{|c|}{0.62} &
\multicolumn{2}{|c|}{0.64} &
\multicolumn{2}{|c}{0.654} \\
\hline 
\end{tabular} \\[5pt]
{
\small 
{\sc Note:}
The standard errors for the entire entries 
in this table are reported to be $\le 0.002$.
}
\end{table}

\noindent 
The design of the simulation is as below:
\begin{enumerate}[label=(\roman*)]
\setlength{\itemsep}{-0.5em}
\item 
{\em Ground-up distribution\/},
that is, a model for $W$ given by (\ref{eqn:LNCDF1}): 
$\mbox{LN} \, (w_0 = 1, \theta = 5, \sigma = 3)$.

\item 
{\em Coinsurance rate\/}: 
$c = 1$.

\item 
{\em Truncation and censoring thresholds 
for both variables $Y$ and $Z$\/}: \\
$d = 4$ 
($\approx 10\%$ left-truncation under 
$\mbox{LN} \, (w_0 = 1, \theta = 5, \sigma = 3)$); \\
$u = 2 \times 10^5$
($\approx 1\%$ right-censoring under 
$\mbox{LN} \, (w_0 = 1, \theta = 5, \sigma = 3)$); \\
$u = 2.4 \times 10^4$
($\approx 4.5\%$ right-censoring under 
$\mbox{LN} \, (w_0 = 1, \theta = 5, \sigma = 3)$).

\item 
{\em Estimators of $\theta$ and $\sigma$\/}: 
\begin{itemize}
\item 
MLE (no trimming on either tail) 

\item 
MTM with different left and right trimming 
proportions $a$ and $b$, respectively, 
satisfying 
\begin{enumerate}[label=(\alph*)]
	\setlength{\itemsep}{-0.5em}
	\item
	$
	0 
	\le 
	a 
	<
	1 - b 
	\le 
	s^{*}
	\le 
	1,
	\
	\mbox{for payment-per-payment variable $Y$}; 
	$ 
	
	\item 
	$
	0 
	\le 
	F_{X}(t)
	\le
	a
	<
	1-b 
	\le 
	F_{X}(T)
	\le 
	1,
	\ 
	\mbox{for payment-per-loss variable $Z$}.
	$
\end{enumerate}
\end{itemize}

\item {\em Sample size\/}: 
$n = 100, \, 250, \, 500, \, 1000$.
\end{enumerate}

\begin{table}[hbt!]
\caption{
Lognormal payment-per-loss 
actuarial loss scenario,
$LN(w_{0}=1,\theta=5,\sigma=3)$ with $d = 4$
and two selected values of right-censoring point $u$.
The entries are mean values 
based on 10,000 samples.
}
\label{table:PPZ_SimStudy2}
\centering
\begin{tabular}{c|cc|cc|cc|cc|cc|cc}
\cline{2-13}
{} & 
\multicolumn{2}{c|}{Proportion} & 
\multicolumn{2}{|c|}{$n=100$} & 
\multicolumn{2}{|c|}{$n=250$} & 
\multicolumn{2}{|c|}{$n=500$} &
\multicolumn{2}{|c|}{$n=1000$} &
\multicolumn{2}{|c}{$n \to \infty$} \\
\cline{2-13} 
& & & & & & & & & & & \\[-2.50ex]
{} & 
$a$ & $b$ & 
$\widehat{\theta}/\theta$ & 
$\widehat{\sigma}/\sigma$ &
$\widehat{\theta}/\theta$ & 
$\widehat{\sigma}/\sigma$ &
$\widehat{\theta}/\theta$ & 
$\widehat{\sigma}/\sigma$ &
$\widehat{\theta}/\theta$ & 
$\widehat{\sigma}/\sigma$ &
$\widehat{\theta}/\theta$ & 
$\widehat{\sigma}/\sigma$ \\
\hline\hline 
\multicolumn{1}{c|}{} &
\multicolumn{12}{c}{} \\[-2.50ex]
\multirow{16}{*}
{\rotatebox{90}{$u=2 \times 10^{5}$}} &
\multicolumn{12}{c}
{Mean values of
	$\widehat{\theta}/\theta$ 
	and
	$\widehat{\sigma}/\sigma$.} \\
\cline{2-13} 
{} & 
\multicolumn{2}{|c|}{MLE} &
1.00 & 1.00 & 
1.00 & 1.00 & 
1.00 & 1.00 & 
1.00 & 1.00 & 
1 & 1 \\
{} & 
0.10 & 0.10 &
1.00 & 1.00 & 
1.00 & 1.00 & 
1.00 & 1.00 & 
1.00 & 1.00 &
1 & 1 \\
{} & 
0.15 & 0.15 &
1.00 & 1.00 & 
1.00 & 1.01 & 
1.00 & 1.00 & 
1.00 & 1.00 &
1 & 1 \\
{} & 
0.25 & 0.25 &
1.00 & 1.01 & 
1.00 & 1.01 &
1.00 & 1.00 &
1.00 & 1.00 &
1 & 1 \\
\cline{2-13}
{} &
0.10 & 0.05 &
1.00 & 1.01 & 
1.00 & 1.00 &
1.00 & 1.00 &
1.00 & 1.00 &
1 & 1 \\
{} &
0.10 & 0.15 &
1.00 & 1.01 & 
1.00 & 1.00 &
1.00 & 1.00 &
1.00 & 1.00 &
1 & 1 \\
{} &
0.10 & 0.25 &
1.00 & 1.01 & 
1.00 & 1.01 &
1.00 & 1.00 &
1.00 & 1.00 &
1 & 1 \\
\cline{2-13}
\multicolumn{1}{c|}{} &
\multicolumn{10}{c}{} \\[-2.50ex]
\multicolumn{1}{c|}{} &
\multicolumn{12}{c}
{Finite-sample efficiencies of MTMs relative to MLEs.} \\
\cline{2-13}
{} & 
\multicolumn{2}{|c|}{MLE} &
\multicolumn{2}{|c}{0.97} & 
\multicolumn{2}{|c}{1.01} &
\multicolumn{2}{|c}{1.01} &
\multicolumn{2}{|c}{0.99} &
\multicolumn{2}{|c}{1} \\
{} & 
0.10 & 0.10 &
\multicolumn{2}{|c|}{0.89} & 
\multicolumn{2}{|c|}{0.86} &
\multicolumn{2}{|c|}{0.88} &
\multicolumn{2}{|c|}{0.85} &
\multicolumn{2}{|c}{0.844} \\
{} & 
0.15 & 0.15 &
\multicolumn{2}{|c|}{0.75} & 
\multicolumn{2}{|c|}{0.73} &
\multicolumn{2}{|c|}{0.76} &
\multicolumn{2}{|c|}{0.74} &
\multicolumn{2}{|c}{0.742} \\
{} & 
0.25 & 0.25 &
\multicolumn{2}{|c}{0.55} & 
\multicolumn{2}{|c|}{0.55} &
\multicolumn{2}{|c|}{0.57} &
\multicolumn{2}{|c|}{0.55} &
\multicolumn{2}{|c}{0.556} \\
\cline{2-13}
{} & 
0.10 & 0.05 &
\multicolumn{2}{|c|}{0.94} & 
\multicolumn{2}{|c|}{0.91} &
\multicolumn{2}{|c|}{0.94} &
\multicolumn{2}{|c|}{0.90} &
\multicolumn{2}{|c}{0.933} \\
{} & 
0.10 & 0.15 &
\multicolumn{2}{|c|}{0.84} & 
\multicolumn{2}{|c|}{0.81} &
\multicolumn{2}{|c|}{0.83} &
\multicolumn{2}{|c|}{0.80} &
\multicolumn{2}{|c}{0.822} \\
{} & 
0.10 & 0.25 &
\multicolumn{2}{|c|}{0.74} & 
\multicolumn{2}{|c|}{0.71} &
\multicolumn{2}{|c|}{0.72} &
\multicolumn{2}{|c|}{0.68} &
\multicolumn{2}{|c}{0.695} \\
\hline\hline 
\multicolumn{1}{c|}{} &
\multicolumn{10}{c}{} \\[-2.50ex]
\multirow{16}{*}
{\rotatebox{90}{$u=2.4 \times 10^{4}$}} &
\multicolumn{12}{c}
{Mean values of
	$\widehat{\theta}/\theta$ 
	and
	$\widehat{\sigma}/\sigma$.} \\
\cline{2-13} 
{} & 
\multicolumn{2}{|c|}{MLE} &
1.00 & 1.00 &  
1.00 & 1.00 & 
1.00 & 1.00 & 
1.00 & 1.00 & 
1 & 1 \\
{} & 
0.10 & 0.10 &
1.00 & 1.00 & 
1.00 & 1.00 &
1.00 & 1.00 &
1.00 & 1.00 & 
1 & 1 \\
{} & 
0.15 & 0.15 &
1.00 & 1.00 & 
1.00 & 1.01 &
1.00 & 1.00 &
1.00 & 1.00 & 
1 & 1 \\
{} & 
0.25 & 0.25 &
1.00 & 1.00 & 
1.00 & 1.01 &
1.00 & 1.00 &
1.00 & 1.00 & 
1 & 1 \\
\cline{2-13}
{} & 
0.10 & 0.15 &
1.00 & 1.00 & 
1.00 & 1.00 &
1.00 & 1.00 &
1.00 & 1.00 & 
1 & 1 \\
{} & 
0.15 & 0.10 &
1.00 & 1.00 & 
1.00 & 1.01 &
1.00 & 1.00 &
1.00 & 1.00 & 
1 & 1 \\
{} & 
0.20 & 0.25 &
1.00 & 1.00 & 
1.00 & 1.01 &
1.00 & 1.00 &
1.00 & 1.00 & 
1 & 1 \\
\cline{2-13}
\multicolumn{1}{c|}{} &
\multicolumn{10}{c}{} \\[-2.50ex]
\multicolumn{1}{c|}{} &
\multicolumn{12}{c}
{Finite-sample efficiencies of MTMs relative to MLEs.} \\
\cline{2-13}
{} & 
\multicolumn{2}{|c|}{MLE} &
\multicolumn{2}{|c|}{0.98} & 
\multicolumn{2}{|c|}{0.99} &
\multicolumn{2}{|c|}{0.99} &
\multicolumn{2}{|c|}{1.00} &
\multicolumn{2}{|c}{1} \\
{} & 
0.10 & 0.10 &
\multicolumn{2}{|c|}{0.92} &
\multicolumn{2}{|c|}{0.90} &
\multicolumn{2}{|c|}{0.89} &
\multicolumn{2}{|c|}{0.88} &
\multicolumn{2}{|c}{0.876} \\
{} & 
0.15 & 0.15 &
\multicolumn{2}{|c|}{0.78} &
\multicolumn{2}{|c|}{0.76} &
\multicolumn{2}{|c|}{0.77} &
\multicolumn{2}{|c|}{0.76} &
\multicolumn{2}{|c}{0.770} \\
{} & 
0.25 & 0.25 &
\multicolumn{2}{|c|}{0.58} & 
\multicolumn{2}{|c|}{0.57} &
\multicolumn{2}{|c|}{0.58} &
\multicolumn{2}{|c|}{0.57} &
\multicolumn{2}{|c}{0.577} \\
\cline{2-13}
{} & 
0.10 & 0.15 &
\multicolumn{2}{|c|}{0.87} &
\multicolumn{2}{|c|}{0.85} &
\multicolumn{2}{|c|}{0.84} &
\multicolumn{2}{|c|}{0.83} &
\multicolumn{2}{|c}{0.822} \\
{} & 
0.15 & 0.25 &
\multicolumn{2}{|c|}{0.68} &
\multicolumn{2}{|c|}{0.67} &
\multicolumn{2}{|c|}{0.67} &
\multicolumn{2}{|c|}{0.67} &
\multicolumn{2}{|c}{0.671} \\
{} & 
0.20 & 0.25 &
\multicolumn{2}{|c|}{0.63} & 
\multicolumn{2}{|c|}{0.62} &
\multicolumn{2}{|c|}{0.62} &
\multicolumn{2}{|c|}{0.62} &
\multicolumn{2}{|c}{0.623} \\
\hline 
\end{tabular} \\[5pt]
{
\small 
{\sc Note:}
The standard errors for the entire entries 
in this table are reported to be $\le 0.001$.
}
\end{table}

For a selected random variable ($Y$ or $Z$),
we generate $10^{4}$ samples of a specified 
size $n$ using the corresponding 
quantile function 
((\ref{p1qf}) for $Y$ and (\ref{p2qf}) for $Z$).
For each sample, depending on 
the underlying scenario, we estimate the parameters 
$\theta$ and $\sigma$ using 
MLE and MTM from both loss variables $Y$ and $Z$.

Simulation results are recorded in Tables 
\ref{table:PPY_SimStudy2}
(for payment-per-payment variable $Y$)
and 
\ref{table:PPZ_SimStudy2}
(for payment-per-loss variable $Z$).
The entries are mean values 
based on $10^4$ samples.  
The columns corresponding to 
$n \rightarrow \infty$ 
represent analytic 
results 
and are found in
Section \ref{sec:MTM}, not from simulations.

As can be seen from Table 
\ref{table:PPY_SimStudy2}
that for $n = 100, \ 250$,
both estimators $\widehat{\theta}$
and $\widehat{\sigma}$ successfully estimate
their corresponding parameter $\theta$ and 
$\sigma$, respectively, with less than
$\pm 2\%$ of the relative bias.
The more important observation is that
they practically become unbiased as soon
as $n \ge 500$.
The situation is even better in Table
\ref{table:PPZ_SimStudy2}.
That is, both estimators $\widehat{\theta}$
and $\widehat{\sigma}$ are practically
unbiased as soon as $n \ge 100$.
On the other hand, as seen in Tables
\ref{table:PPY_SimStudy2} and
\ref{table:PPZ_SimStudy2},
the finite REs are
obviously biased (as expected)
for small sample sizes
and converge slower to their corresponding
ARE levels. But what is clear is that
all those REs are asymptotically unbiased.

\section{Numerical Examples}
\label{sec:numericalExamples}

\begin{figure}[b!]
\centering
\includegraphics[width=1.10\textwidth]
{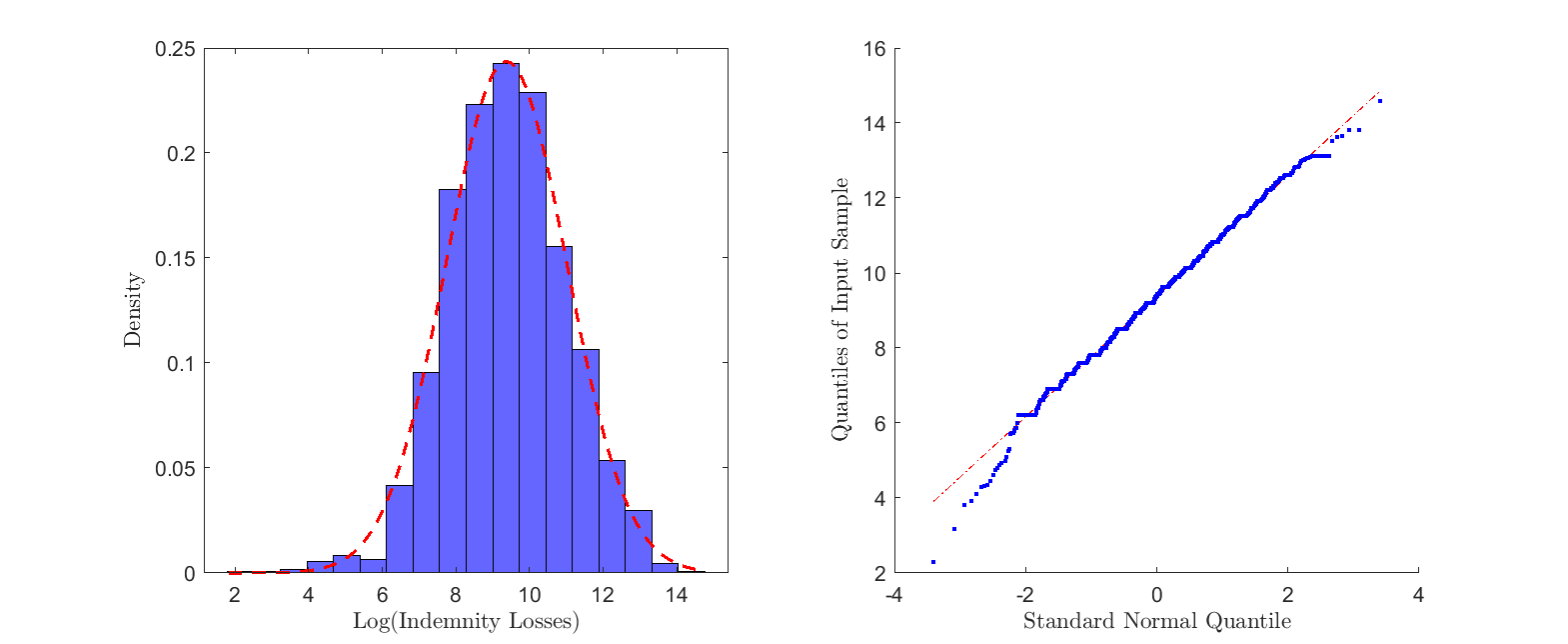}
\caption{Histogram of log-transformed indemnity losses
with the fitted normal density (left panel) and the 
corresponding quantile-quantile plot (right panel).}
\label{fig:IndemnityDiagnostics1}
\end{figure}

This section is to observe the performance of the 
estimation methods developed in the previous sections.
For the illustration purpose only, we consider the 
1500 US indemnity losses 
which is widely studied in actuarial literature,
see, for example, 
\cite{MR1988432,MR3836650,mmm20}.
We then transform the data to fit 
the scenarios of payment-per-payment random variable $Y$
and payment-per-loss random variable $Z$. 
That is, from an insurer's prospective, we consider
an insurance benefit equal to the amount by which a 
loss exceeds US\$500.00
(deductible, $d=500)$ but with a maximum benefit 
of US\$100,000.00 (policy limit, $u=10^5)$.
Further, without loss of generality, since the asymptotic 
variances of all the estimators investigated in this paper
do not depend on the coinsurance factor $c$, we simply
assume that $c \equiv 1$. 
We consider this example purely for illustrative 
purpose  and a preliminary diagnostics,
see Figure \ref{fig:IndemnityDiagnostics1},
shows that
the lognormal distribution provides a satisfactory,
though not perfect, fit to the indemnity data.
Moreover, the complete ground-up data set 
(though the data set contains some right-censored observations) 
is identified to fit 
$LN(w_{0}=0,\theta,\sigma)$, 
\label{LN:Example1}
see, for example, 
\cite{,MR3836650,mmm20},
distribution. 
The Kolmogorov-Smirnov (KS) test statistics
\citep[see, e.g.,][\S15.4.1, p. 360]{MR3890025}
is computed to be 0.0266 and with the significance
level of 5\% the corresponding critical value is 
0.0351. 
Therefore, the lognormal model is a plausible
model for the indemnity loss data set at 5\%
significance level.
The specified insurance contract data transformation
is summarized in Table \ref{table:Data_Transformation}.

\begin{table}[t!]
\centering
\caption{
Payment-per-payment, $Y$ and payment-per-loss, $Z$
transformation scenarios for the US indemnity losses.
}
\label{table:Data_Transformation}
\begin{tabular}{r|c|c|c|c|ccccc}
\hline\hline
\multirow{2}{*}{Scenario} & 
\multicolumn{2}{c|}{Left truncation} & 
\multicolumn{2}{c|}{Right censorship} & 
\multirow{2}{*}{$n_0$} & 
\multirow{2}{*}{$n_{1}$} & 
\multirow{2}{*}{$n_{2}$} & 
\multirow{2}{*}{$n$} & 
\multirow{2}{*}{$R$}\\
\cline{2-5}
{} & 
LN -- $d$ & Normal -- $t$ &
LN -- $u$ & Normal -- $T$ &
{} & {} & {} & {} & {} \\
\hline 
Payment $Y$ & $500$ & 
$6.2146$ & $10^5$ & 
$11.5129$ & -- & 1299 & 152 & 1451 & 5.2983 \\
\hline 
Payment $Z$ & $500$ & 
$6.2146$ & $10^5$ & 
$11.5129$ & 49 & 1299 & 152 & 1500 & 5.2983 \\
\hline\hline 
\end{tabular} \\
Note: LN stands for LogNormal, 
$t = \log{(d-w_{0})}$, and 
$T = \log{(u-w_{0})}$.
\end{table}

A couple of crucial points are in queue to be mentioned
here before starting real data analysis. 
First, for payment $Y$, the MTM estimators as well
as their asymptotic results are based on the assumption
of $0 \le a < 1-b \le s^{*} \le 1$. 
Initially, we set
\[
\hat{s}_{\mbox{\tiny E}}^{*}
:=
\dfrac{F_{n}(T)-F_{n}(t)}{1-F_{n}(t)}
=
{n_{1}}/{n},
\quad 
\mbox{where $F_{n}$ is the empirical cdf},
\]
and choose $b$ such that 
$1-b \le \hat{s}_{\mbox{\tiny E}}^{*}$.
But after estimating 
$\widehat{\theta}_{\mbox{\tiny y,MTM}}$ and 
$\widehat{\sigma}_{\mbox{\tiny y,MTM}}$,
it could be the case that 
\[
\hat{s}_{\mbox{\tiny P}}^{*}
:=
\dfrac{\widehat{F}_{X}(T)-\widehat{F}_{X}(t)}
{1-\widehat{F}_{X}(t)}
< 1-b,
\quad 
\mbox{where $\widehat{F}_{X}$ is the estimated parametric cdf},
\]
which is unsatisfactory for statistical inferences as 
it violates the assumption 
(Case 2 from Section \ref{sec:PPY_MTM1}).
Therefore, the value of $b$ should be chosen dynamically
such that  
$1-b \le 
\min\{
\hat{s}_{\mbox{\tiny E}}^{*},
\hat{s}_{\mbox{\tiny P}}^{*}
\}.
$
Similarly, for payment $Z$, the left and right trimming proportions
$a$ and $b$, respectively, should be chosen dynamically as:
\[
\max \{F_{n}(t),\widehat{F}_{X}(t)\}
\le 
a
\quad \mbox{and} \quad 
1-b 
\le
\min \{F_{n}(T),\widehat{F}_{X}(T)\}.
\]
The trimming proportions in Tables \ref{table:estTablePymtY}
and \ref{table:estTablePymtZ} are chosen accordingly.

\begin{figure}[b!]
\centering
\includegraphics[width=0.98\textwidth]{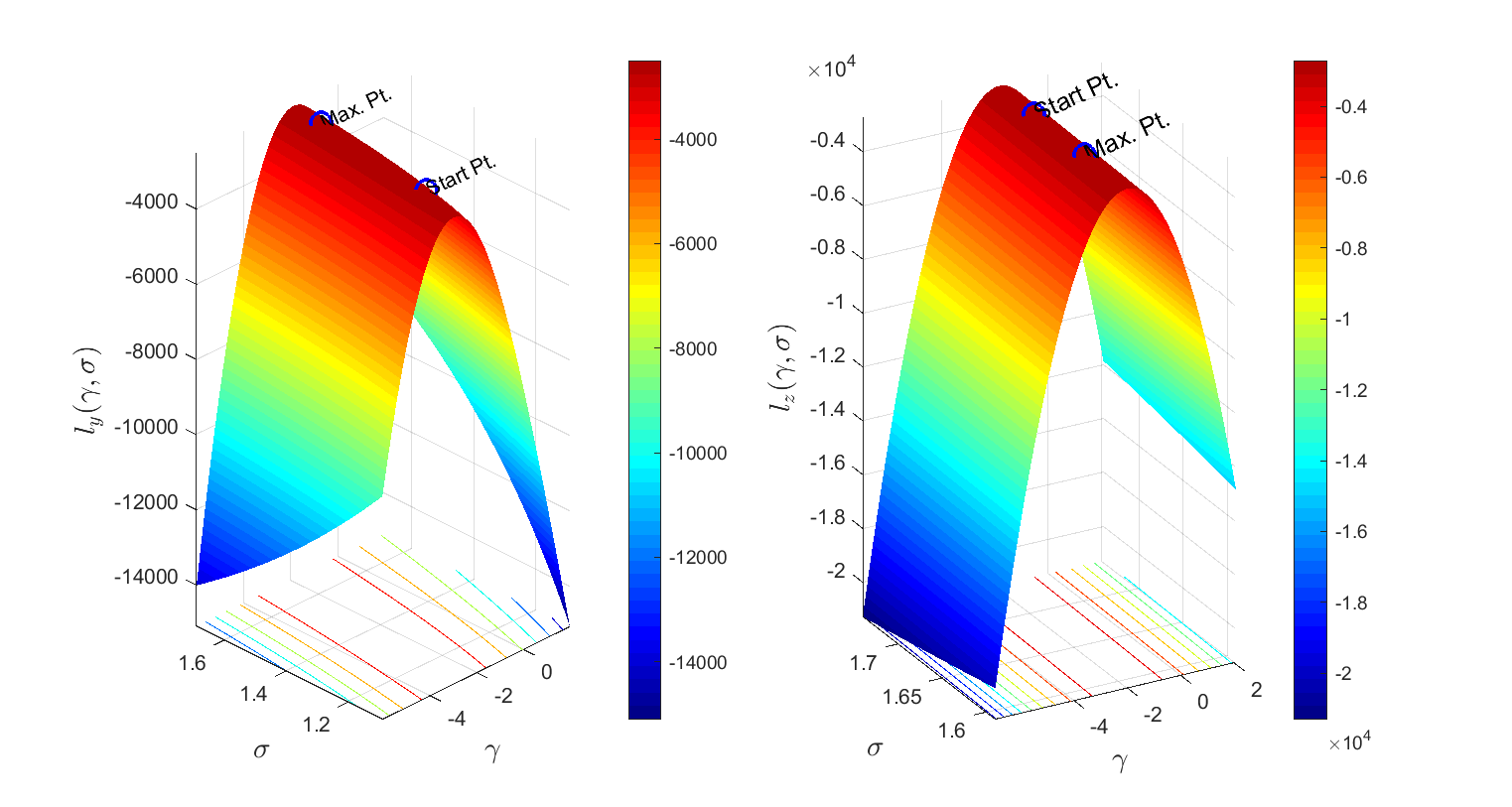}
\caption{Log-likelihood surfaces for payment-per-payment
(left panel) and payment-per-loss (right panel) data 
scenarios given by Equations (\ref{eqn:PPYMLELogLik1}) and 
(\ref{eqn:PPZMLELogLik1}), respectively.}
\label{fig:likelihoodSurf}
\end{figure}

For payment-per-payment data scenario, the MLE 
system of Equations (\ref{eqn:PPNormalMLEeqn2}) 
takes the form:
\begin{align}
\label{eqn:PPNormalMLEeqn2Data}
\left\{
\begin{array}{rrr}
\sigma\left(\Omega_{y,1}-\Omega_{y,2}-\gamma\right) - 2.9762
& = & 0, \\[10pt]
\sigma^{2}\left(1-\gamma(\Omega_{y,1}-\Omega_{y,2}-\gamma)
-\frac{5.2983 \Omega_{y,2}}{\sigma}\right) -  10.3394 
& = & 0,
\end{array} \right.
\end{align}
with the first approximations:
$(\gamma_{\mbox{\tiny start}}, 
\sigma_{\mbox{\tiny start}})
=
(-2.4453,1.2171)$, denoted by "Start Pt."
on the corresponding log-likelihood surface given by 
Figure \ref{fig:likelihoodSurf} (left panel),
but note that $\Omega_{y,1}$ and $\Omega_{y,2}$
also depend on $\gamma$ and $\sigma$.
The final MLE solution 
$
\left(
\widehat{\gamma}_{\mbox{\tiny y,MLE}}, \widehat{\sigma}_{\mbox{\tiny y,MLE}}
\right)
$
of the system (\ref{eqn:PPNormalMLEeqn2Data})
is identified with "Max. Pt." in Figure \ref{fig:likelihoodSurf} (left panel).
Similarly, with different left and right trimming 
proportions, the system of MTM equations given by
(\ref{eqn:Normal_MTMPPY_Eqns}) can be solved with their
respective first approximations given by 
(\ref{eqn:Normal_MTMPPY_EqnsStart}).
The estimated values with their corresponding asymptotic
$95\%$ confidence intervals are displayed in Table 
\ref{table:estTablePymtY}. 
The confidence intervals for $\sigma$ provided are
log-transformed confidence intervals 
\cite[see, e.g.,][p. 315-316]{MR3890025}
since some of the linear confidence intervals for 
$\sigma$ include $0$ which are unsatisfactory as
$\sigma > 0$. 
For completely observed ground-up lognormal severity 
data,  
$
\mbox{ARE}
\left(
\left(
\widehat{\theta}_{\mbox{\tiny MTM}},
\widehat{\sigma}_{\mbox{\tiny MTM}}
\right),
\left(
\widehat{\theta}_{\mbox{\tiny MLE}},
\widehat{\sigma}_{\mbox{\tiny MLE}}
\right)
\right)
$
does not depend on any unknown parameters to be
estimated \citep{MR2497558}. 
But for payment $Y$ and payment $Z$ data scenarios,
the corresponding AREs, respectively, given by 
(\ref{eqn:PPY_MTM_MLE_ARE}) and (\ref{eqn:PPZ_MTM_MLE_ARE})
depend on the unknown parameters $\theta$ and $\sigma$.
Therefore, for comparison purpose, in Tables 
\ref{table:estTablePymtY} and \ref{table:estTablePymtZ},
the $\widehat{\mbox{ARE}}$ column is computed
as a bivariate function of MLE estimated values 
$\widehat{\theta}_{\mbox{\tiny MLE}}$ and 
$\widehat{\sigma}_{\mbox{\tiny MLE}}$.
As seen in Table \ref{table:estTablePymtY}, 
the estimated values of $\theta$ and $\sigma$ 
via MTM are seem to underestimate the MLE 
estimated values 
$
\left( 
\widehat{\theta}_{\mbox{\tiny MLE}},
\widehat{\sigma}_{\mbox{\tiny MLE}}
\right)
=
\left( 
9.43, 1.59
\right) 
$
but it is important to note that 
all the MTM estimated values of $\theta$
and $\sigma$ are close enough to the 
corresponding MLE estimator even after removing
some of the point mass at the right-censored point.
It is also noticeable that with trimming pair 
$
(a,b) 
=
(100/1451,300/1451),
$
the estimated values of $\theta$ and $\sigma$
are almost equal to the corresponding MLE
estimators but still maintaining the 
efficiency about of 79\%. 
These all demonstrate the robustness of MTM
estimators.

\begin{table}[t]
\centering
\caption{MLE and MTM estimators of $\theta$ and $\sigma$ 
with their corresponding asymptotic confidence intervals
and estimated values of AREs for payment $Y$ data scenario.}
\label{table:estTablePymtY}
\begin{tabular}{r|c|c|cc|cc|cc|c|cc}
\hline
\multicolumn{3}{c|}{\multirow{2}{*}{Estimators}} &
\multirow{2}{*}{$\widehat{\theta}$} & 
\multirow{2}{*}{$\widehat{\sigma}$} & 
\multicolumn{2}{|c|}{$\hat{s}^{*}$} & 
\multicolumn{2}{|c|}{95\% CI for} & 
\multirow{2}{*}{$\widehat{\mbox{ARE}}$} & 
\multicolumn{2}{|c}
{\multirow{1}{*}{KS Test}} \\
\cline{6-9}
\multicolumn{2}{c}{} & {} & {} & {} & 
$\hat{s}_{\mbox{\tiny E}}^{*}$ &
$\hat{s}_{\mbox{\tiny P}}^{*}$ &
$\theta$ & $\sigma$ & {} & 
$D_{y}$ & $h_{y}$ \\
\hline\hline
\multicolumn{3}{c|}{MLE} & 
9.43 & 1.59 & -- & -- & $(9.34,9.52)$ & $(1.52,1.67)$ &
1.00 & 
.032 & 0 \\
\cline{1-12}
\multirow{9}{*}{\rotatebox{90}{MTM}} & $a$ & $b$ &
\multicolumn{9}{c}
{Condition required: 
	$
	1-b 
	\le 
	\min\{
	\hat{s}_{\mbox{\tiny E}}^{*},\hat{s}_{\mbox{\tiny P}}^{*}
	\}.
	$} \\
\hline 
{} & 0.00 & 150/1451 & 
9.42 & 1.56 & 0.90 & 0.91 & $(9.34,9.51)$ & $(1.49,1.65)$ &
0.94 & 
.034 & 0 \\
{} & 0.00 & 200/1451 & 
9.42 & 1.55 & 0.90 & 0.91 & $(9.33,9.51)$ & $(1.47,1.64)$ &
0.89 & 
.034 & 0 \\
{} & 0.00 & 300/1451 &
9.42 & 1.54 & 0.90 & 0.91 & $(9.33,9.50)$ & $(1.45,1.63)$ &
0.80 &
.034 & 0 \\
{} & 0.00 & 700/1451 &
9.37 & 1.47 & 0.90 & 0.93 & $(9.27,9.47)$ & $(1.35,1.59)$ &
0.48 &
.043 & 1 \\
\cline{2-12}
{} & 10/1451 & 150/1451 & 
9.42 & 1.57 & 0.90 & 0.91 & $(9.33,9.51)$ & $(1.49,1.65)$ &
0.94 & 
.033 & 0\\
{} & 50/1451 & 200/1451 &
9.41 & 1.59 & 0.90 & 0.91 & $(9.32,9.50)$ & $(1.50,1.67)$ &
0.89 &
.030  & 0 \\
{} & 100/1451 & 300/1451 & 
9.40 & 1.59 & 0.90 & 0.90 & $(9.31,9.50)$ & $(1.50,1.69)$ &
0.79 & 
.028 & 0 \\
{} & 650/1451 & 650/1451 & 
9.26 & 2.09 & 0.90 & 0.85 & $(8.96,9.56)$ & $(1.56,2.81)$ &
0.24 & 
.064 & 1 \\
\hline
\end{tabular} \\[5pt]
{\sc Note:}
{\scriptsize 
For the KS Test column, 
$D_{y}$ stands for Kolmogorov-Smirnov
test statistics and $h_{y} \in \{0,1\}$
stands for the corresponding decision.
$h_{y} = 0$ means that the assumed model
is plausible and $h_{y} = 1$ means that 
the model is rejected.
}
\end{table}

In order to conduct hypothesis test
for the various fitted models, we use
the KS test statistic.
The KS test statistic for 
payments $Y$ data set is defined as
\citep[see, e.g.,][\S15.4.1, p. 360]{MR3890025}:
\begin{align}
\label{eqn:KS_PPY_Defn}
D_{y}
& :=
\max_{0 < y
\leq 
c(u-d)}
\left|F_{n}(y)-\widehat{F}_{Y}(y)\right|,
\end{align}
where $\widehat{F}_{Y}$ is the estimated 
cdf given by 
\eqref{p1cdf}.
For each fitted model, the corresponding
KS test statistics $(D_{y})$ and the decision $(h_{y})$
of the hypothesis test are given in the 
last two columns of Table 
\ref{table:estTablePymtY}
where $h_{y} = 0$ indicates that the lognormal model
is plausible and $h_{y} = 1$ means that the 
the lognormal model is rejected at the 
significance level of $5\%$.
Only one model fitted  with 
$
(a,b) 
=
\left( 
650/1451,650/1451
\right)
\approx
(.45,.45)
$
is rejected which is close to median-type estimator.

Similarly, for payment-per-loss data scenario, 
the MLE system of Equations (\ref{eqn:censoredNormalMLEeqn2}) 
takes the form:
\begin{align}
\label{eqn:censoredNormalMLEeqn2Data}
\left\{
\begin{array}{rrr}
\sigma\left(\Omega_{z,1}-\Omega_{z,2}-\gamma\right) - 2.9762 
& = & 0, \\[10pt]
\sigma^{2}\left(1-\gamma(\Omega_{z,1}-\Omega_{z,2}-\gamma)
-\frac{5.2983 \Omega_{z,2}}{\sigma}\right) -  10.3394 
& = & 0,
\end{array} \right.
\end{align}
with the first approximations:
$(\gamma_{\mbox{\tiny start}}, 
\sigma_{\mbox{\tiny start}})
=
(-1.8430,1.6998)$.
The starting point and the final MLE 
solution, respectively, are represented by "Start Pt." 
and "Max. Pt." on the log-likelihood surface $l_{z}(\gamma,\sigma)$
given by Figure \ref{fig:likelihoodSurf} (right panel).
Including the corresponding MTM calculation,
a summary 
statistics is displayed in Table \ref{table:estTablePymtZ}.
In this case, as expected all the MTM
estimators of $\theta$ are same which
clearly demonstrate the stability of 
the MTM methodology.
On the other hand, highly robust MTM estimated value
of $\widehat\sigma = 2.36$ with 
$
(a,b) 
=
(700/1500,700/1500)
\approx 
(.47,.47)
$
seems to be more volatile
and loosing about 84\% of 
asymptotic efficiency. 

\begin{table}[t!]
\centering
\caption{MLE and MTM estimators of $\theta$ and $\sigma$ 
with their corresponding asymptotic confidence intervals
and estimated values of AREs for payment $Z$ data scenario.}
\label{table:estTablePymtZ}
\begin{tabular}{r|c|c|cc|cc|cc|c|cc}
\hline
\multicolumn{3}{c|}{\multirow{2}{*}{Estimators}} &
\multirow{2}{*}{$\widehat{\theta}$} & 
\multirow{2}{*}{$\widehat{\sigma}$} & 
\multirow{2}{*}{\small$\widehat{F}_{X}(t)$} & 
\multirow{2}{*}{\small$\widehat{F}_{X}(T)$} & 
\multicolumn{2}{|c|}{95\% CI for} & 
\multirow{2}{*}{$\small\widehat{\mbox{ARE}}$} & 
\multicolumn{2}{|c}{KS Test}\\
\cline{8-9}
\multicolumn{2}{c}{} & {} & {} & {} & {} & {} &
$\theta$ & $\sigma$ & {} &
$D_{z}$ & $h_{z}$ \\
\hline\hline
\multicolumn{3}{c|}{MLE} & 
9.39 & 1.64 & -- & -- & $(9.30,9.47)$ & $(1.58,1.71)$ &
1.00 & 
.027 & 0 \\
\hline 
\multirow{10}{*}{\rotatebox{90}{MTM}} & 
\multirow{2}{*}{$a$} & 
\multirow{2}{*}{$b$} &
\multicolumn{9}{c}{Condition required:} \\
{} & {} & {} & 
\multicolumn{9}{c}
{
	$
	\max \{F_{n}(t),\widehat{F}_{X}(t)\} \le a
	\quad \mbox{and} \quad 
	1-b \le \min \{F_{n}(T),\widehat{F}_{X}(T)\}.
	$
} \\
\cline{2-12}
{} & 75/1500 & 150/1500 & 
9.38 & 1.62 & 0.03 & 0.91 & $(9.30,9.47)$ & $(1.55,1.69)$ &
0.92 & 
.027 & 0 \\
{} & 75/1500 & 225/1500 & 
9.38 & 1.61 & 0.02 & 0.91 & $(9.30,9.47)$ & $(1.54,1.69)$ &
0.86 & 
.027 & 0 \\
{} & 75/1500 & 375/1500 & 
9.38 & 1.60 & 0.02 & 0.91 & $(9.29,9.46)$ & $(1.53,1.69)$ &
0.76 & 
.027 & 0 \\
{} & 75/1500 & 750/1500 & 
9.36 & 1.59 & 0.02 & 0.91 & $(9.26,9.47)$ & $(1.49,1.70)$ &
0.52 & 
.028 & 0 \\
\cline{2-12}
{} & 150/1500 & 150/1500 & 
9.38 & 1.63 & 0.03 & 0.90 & $(9.30,9.47)$ & $(1.55,1.70)$ &
0.86 & 
.026 & 0 \\
{} & 225/1500 & 225/1500 & 
9.38 & 1.63 & 0.03 & 0.90 & $(9.29,9.46)$ & $(1.55,1.72)$ &
0.76 & 
.026 & 0 \\
{} &375/1500 & 375/1500 & 
9.38 & 1.61 & 0.02 & 0.91 & $(9.29,9.47)$ & $(1.50,1.71)$ &
0.57 & 
.027 & 0 \\
{} & 700/1500 & 700/1500 & 
9.38 & 2.36 & 0.09 & 0.82 & $(9.23,9.52)$ & $(1.92,2.91)$ &
0.16 & 
.107 & 1 \\
\hline
\end{tabular} \\[5pt]
{\sc Note:}
{\scriptsize 
$
\left( 
F_{n}(t),F_{n}(T) 
\right)
= 
(49/1500,1348/1500)
\approx
(0.03,0.90).
$
For the KS Test column, 
$D_{z}$ stands for Kolmogorov-Smirnov
test statistics and $h_{z} \in \{0,1\}$
stands for the corresponding decision.
$h_{z} = 0$ means that the assumed model
is plausible and $h_{z} = 1$ means that 
the model is rejected.
}
\end{table}  

Further, the KS test statistic for payments $Z$ data set 
is defined by:
\begin{align}
\label{eqn:KS_PPZ_Defn}
D_{z}
& :=
\max_{0 \le z
\leq 
c(u-d)
}
\left|F_{n}(z)-\widehat{F}_{Z}(z)\right|,
\end{align}
where $\widehat{F}_{Z}$ is the estimated 
parametric cdf given by 
\eqref{p2cdf}.
As can be seen in Table \ref{table:estTablePymtZ},
similar to \ref{table:estTablePymtZ},
only one model fitted for payment-per-loss
data scenario with 
$
(a,b) 
= 
(700/1500,700/1500)
\approx 
(.47,.47)
$
is rejected with the significance level
of $5\%$, but this model is loosing about
84\% of the ARE compared to the corresponding MLE model.

In conclusion, a take-home knowledge from 
Tables \ref{table:estTablePymtY} and 
\ref{table:estTablePymtZ} is that since 
the sample data is already either truncated
and/or censored, then a robust estimation procedure
still maintaining high relative efficiency with 
respect to the corresponding MLE is simply to 
eliminate the point masses at the truncated and 
censored values.
For example, 
$
MTM(a = 0, b = 150/1451)
$
and 
$
MTM(a = 10/1451, b = 150/1451)$
from Table \ref{table:estTablePymtY} and 
$MTM(a = 75/1500, b = 150/1500)$ 
from Table \ref{table:estTablePymtZ}.

\section{Summary and Future Direction}
\label{sec:Conclusion}

In this paper, 
we have developed two estimation procedures -- 
{\em maximum likelihood} and a dynamic 
MTM for 
lognormal insurance payment-per-payment and 
payment-per-loss loss severity data.
A series of theoretical results about estimators'
existence and asymptotic normality are established.
As seen in Tables \ref{table:PPY_MLE_MTM_ARE}
and \ref{table:PPZ_MLE_MTM_ARE},
the dynamic MTM estimators sacrifice little
asymptotic efficiency with respect to MLE 
but are more robust as well as computationally
more efficient than MLE if properly implemented, 
see, for example, (\ref{eqn:PPY_MTM_Estimators1})
and Note \ref{note:PPY_MTM_Symplified1},
and these properties are highly desirable in practice.
Finally, the developed estimators are implemented
on a real-life data set to analyze the 1500 US indemnity
losses which is widely studied in the actuarial literature
\citep[see, e.g.,][]{MR1988432,mmm20}, and it is found
that most of the fitted models are plausible for this data set.

The results of this paper motivate open problems 
and generate several ideas for further research. 
First, this paper is specifically focused on 
lognormal insurance payment severity models, 
but they could be extended to more general (log) 
location-scale and exponential dispersion families.
Second, several contaminated loss severity models are proposed 
in the literature as mentioned in Section \ref{sec:Introduction}
Introduction,
so it could even produce a better model still maintaining 
a reasonable balance between efficiency and robustness
if one implements the dynamic MTM procedures for those
mixture models, but it is recommended to design the MTM 
for the complete mixture models first.
Third, in this paper we only discussed about model-fitting 
procedures and the corresponding finite sample performance,
but the ultimate goal of model fitting is 
to apply the fitted models in decision-making mechanism.
Therefore, it is yet to measure how the estimators developed 
in this paper act in calculating actuarial pure premium 
as well as in different risk analysis in practice. 

Finally, 
it is important to emphasize again that the motivation to 
implement the MTM approach rather than MLE even for truncated 
and/or censored loss data sets is to simultaneously remove 
partial point masses assigned by MLE at the truncated and/or
censored data points
(that is, achieving a desired degree of robustness) 
and maintaining a high degree of ARE, too. 
Further, 
the dynamic MTM procedures designed in this paper
can only be implemented if a close fit in one or both tails
of the assumed underlying distribution is not desired, otherwise
extreme value modeling is recommended. 

\newpage 

\baselineskip 4.80mm
\setlength{\bibsep}{5.0pt plus 0.00ex}
\bibliography{ArXivMain}
\thispagestyle{plain}
\pagestyle{plain} 
\thispagestyle{plain}

\newpage 

\begin{appendices}

\section{Auxiliary Results}
\label{sec:Appendix}

Define 
$\bar{a} := 1-a$,
$\bar{b} := 1-b$, and
$\tau := 1-a-b$.
\begin{enumerate}[label=\Roman*.]

\item 
The expressions for 
$c_{k}, \ 1 \le k \le 4$ 
mentioned in Section \ref{sec:PPZ_MTM1}
are given by
\begin{align*}
	c_{k} 
	& \equiv 
	c_{k}(\Phi,a,b)
	=
	\dfrac{1}{\tau} 
	\int_{a}^{\bar{b}} 
	\left[ 
	\Phi^{-1}(s)
	\right]^{k} \, ds.
\end{align*}

\item 
The expressions for 
$c_{k}^{*}, \ 1 \le k \le 3$
used in Section \ref{sec:PPZ_MTM1}
are given as below:
\begin{align*}
	c_{1}^{*} 
	& = 
	\left( 
	\dfrac{1}{\tau}
	\right)^{2} 
	\int_{a}^{\bar{b}} 
	\int_{a}^{\bar{b}} 
	{\frac{\min\:\{u,v\}-uv}
		{\phi\left[\Phi^{-1}\left(v\right)\right]
			\phi\left[\Phi^{-1}\left(u\right)\right]}} 
	\, dv \, du
	\\[5pt] 
	& = 
	\left( 
	\dfrac{1}{\tau}
	\right)^{2} 
	\left\{
	a \bar{a} \left[ \Phi^{-1}(a)\right]^{2}
	+ 
	b\bar{b}\left[ \Phi^{-1}(\bar{b})\right]^{2}
	-
	2ab \Phi^{-1}(a) \Phi^{-1}(\bar{b})
	\right. \\
	& \quad 
	\left. 
	-2\tau c_{1}
	\left[
	a \Phi^{-1}(a) + b \Phi^{-1}(\bar{b})
	\right]
	- 
	\tau^{2}c_{1}^{2} 
	+ 
	\tau c_{2}
	\right\}.
	\\[5pt]
	c_{2}^{*} 
	& = 
	\left( 
	\dfrac{1}{\tau}
	\right)^{2}
	\int_{a}^{\bar{b}} 
	\int_{a}^{\bar{b}} 
	{\frac{\left(\min\:\{u,v\}-uv\right)\Phi^{-1}(u)}
		{\phi\left[\Phi^{-1}\left(v\right)\right]
			\phi\left[\Phi^{-1}\left(u\right)\right]}} 
	\, dv \, du 
	\\[5pt]
	& =
	\dfrac{1}{2\tau^{2}} 
	\left\{ 
	a\bar{a} \left[ \Phi^{-1}(a)\right]^{3}
	+ 
	b\bar{b}\left[ \Phi^{-1}(\bar{b})\right]^{3} 
	-
	ab \Phi^{-1}(a) \Phi^{-1}(\bar{b})
	\left[ 
	\Phi^{-1}(a) + \Phi^{-1}(\bar{b})
	\right] 
	\right. 
	\\[5pt] 
	& \quad 
	\left. 
	-
	\tau c_{1}
	\left[ 
	a \left[ \Phi^{-1}(a)\right]^{2}
	+ 
	b \left[ \Phi^{-1}(\bar{b})\right]^{2}
	\right] 
	-\tau c_{2}
	\left[ 
	a\Phi^{-1}(a) + b\Phi^{-1}(\bar{b}) 
	\right]
	-
	\tau^{2}c_{1}c_{2}
	+ 
	\tau c_{3}
	\right\}.
	\\[5pt]
	c_{3}^{*} 
	& = 
	\left( 
	\dfrac{1}{\tau}
	\right)^{2}
	\int_{a}^{\bar{b}} 
	\int_{a}^{\bar{b}} 
	{\frac{\left(\min\:\{u,v\}-uv\right)
			\Phi^{-1}(v)\Phi^{-1}(u)}
		{\phi\left[\Phi^{-1}\left(v\right)\right]
			\phi\left[\Phi^{-1}\left(u\right)\right]}} 
	\, dv \, du 
	\\[5pt] 
	& = 
	\dfrac{1}{4\tau^{2}} 
	\left\{ 
	a\bar{a} \left[ \Phi^{-1}(a)\right]^{4}
	+ 
	b\bar{b}\left[ \Phi^{-1}(\bar{b})\right]^{4} 
	-
	2ab 
	\left[\Phi^{-1}(a)\right]^{2}
	\left[\Phi^{-1}(\bar{b})\right]^{2}
	\right. 
	\\[5pt] 
	& \quad 
	\left. 
	-
	2\tau c_{2}
	\left[ 
	a \left[ \Phi^{-1}(a)\right]^{2}
	+ 
	b \left[ \Phi^{-1}(\bar{b})\right]^{2}
	\right] 
	-\tau^{2} c_{2}^{2}
	+
	\tau c_{4}
	\right\}.
\end{align*}

\item 
The entries of the variance-covariance matrix 
$\bm{\Sigma}_{y}$ from Section \ref{sec:PPY_MTM1}
are as follows:
\begin{align*}
	c_{y,1}^{*}
	& = 
	\left(
	\frac{\bar{\Phi}(\gamma)}{\tau}
	\right)^{2}
	\int_{a}^{\bar{b}}\int_{a}^{\bar{b}}{\frac{\text{min}\:\{u,v\}-uv}{\phi\left[\Phi^{-1}\left(v+(1-v)\Phi(\gamma)\right)\right]\phi\left[\Phi^{-1}\left(u+(1-u)\Phi(\gamma)\right)\right]}} \, dv \, du. \\[5pt]
	c_{y,2}^{*}
	& = 
	\left(
	\frac{\bar{\Phi}(\gamma)}{\tau}
	\right)^{2}
	\int_{a}^{\bar{b}}\int_{a}^{\bar{b}}{\frac{\left[\text{min}\:\{u,v\}-uv\right]\Phi^{-1}\left(u+(1-u)\Phi(\gamma)\right)}{\phi\left[\Phi^{-1}\left(v+(1-v)\Phi(\gamma)\right)\right]\phi\left[\Phi^{-1}\left(u+(1-u)\Phi(\gamma)\right)\right]}} \, dv \, du. \\[5pt]
	c_{y,3}^{*} 
	& = 
	\left(
	\frac{\bar{\Phi}(\gamma)}{\tau}
	\right)^{2}
	\int_{a}^{\bar{b}}\int_{a}^{\bar{b}} \left[ {\frac{\left[\text{min}\:\{u,v\}-uv\right]\Phi^{-1}\left(v+(1-v)\Phi(\gamma)\right)}{\phi\left[\Phi^{-1}\left(v+(1-v)\Phi(\gamma)\right)\right]}}\right. \\
	& {\qquad \qquad \qquad \qquad \qquad \qquad \qquad \qquad} \times \left. {\frac{\Phi^{-1}\left(u+(1-u)\Phi(\gamma)\right)}{\phi\left[\Phi^{-1}\left(u+(1-u)\Phi(\gamma)\right)\right]}}\right] \, dv \, du.
\end{align*}

\end{enumerate}

\end{appendices}

\end{document}